\documentstyle[12pt,epsf]{article}
\def\ov{\overline}
\def\s2{\frac{1}{\sqrt2}}

\def\beq{\begin{equation}}
\def\eeq{\end{equation}}
\def\beqa{\begin{eqnarray}}
\def\eeqa{\end{eqnarray}}
\def\times{\otimes}
\def\IR{\relax{\rm I\kern-.18em R}}
\def\IP{\relax{\rm I\kern-.18em P}}
\def\IC{\relax\hbox{\kern.25em$\inbar\kern-.3em{\rm C}$}}

\def\cp#1{\relax\ifmmode {\IP\kern-2pt{}_{#1}}\else
$\IP\kern-2pt{}_{#1}$\fi}

\def\im{\hbox{Im\,}}
\def\re{\hbox{Re\,}}

\def\WW{W^\alpha W_\alpha}
\def\lbarl{\lambda^\alpha \lambda_\alpha}
\def\IZ{{\bf Z}}
\def\Avg#1{\left\langle #1 \right\rangle}

\def\d{\partial}

\renewenvironment{thebibliography}[1]
 { \small
   \begin{list}{\arabic{enumi}.}
    {\usecounter{enumi} \setlength{\parsep}{0pt}
     \setlength{\itemsep}{3pt} \settowidth{\labelwidth}{#1.}
     \sloppy
    }}{\end{list}}
 
\begin{document}
\begin{titlepage}
\begin{flushright}
CERN-TH/96-65 \\
hep-th/9603074 \\
\end{flushright}
\vskip 0.3in

\vglue 0.4cm
\begin{center}
 {\LARGE 		Lectures on Superstring Phenomenology
			%\vglue 10pt
			  }
\vglue 1.8cm
{\large 		Fernando Quevedo }
\baselineskip=12pt
\vglue 0.7cm
{\small\em 		Theory Division, CERN 	\\
			CH-1211, Geneva 23\\  
			Switzerland }
\end{center}

\vglue 2.0cm
{\rightskip=3pc
 \leftskip=3pc
 \tenrm\baselineskip=15pt
 \noindent
\begin{center}
{\tenrm \bf Abstract}
\end{center}
\vglue 0.6cm

\normalsize
{\noindent}The phenomenological aspects of string theory are
briefly reviewed. Emphasis is given to the status
of 4D string model building, effective Lagrangians, model
independent results, supersymmetry breaking and
duality symmetries.

\vglue 2.8cm}
\begin{center}
{\em Lectures Given at V Latin American Workshop on}\\
{\em Particles and Fields, Puebla, Mexico, November 1995}
\end{center}
\vskip4.8cm

\begin{flushleft}
CERN-TH/96-65 \\
March 1996 \\
\end{flushleft}
\end{titlepage}

{\large\bf\noindent Introduction }
\normalsize
\vglue 0.4cm
\baselineskip=15pt

We all know that quantum field theory (QFT) is the fundamental stone on which high energy
physics is based
\cite{steve}. This theory 
was developed  in order to make consistent
the general principles of special relativity and quantum mechanics,
both fundamental to the study of elementary particles.
Given the generality of QFT, there are very few general consequences 
we can extract from it. We can only mention: the existence of antiparticles,
the running of coupling constants, the relation between spin 
and statistics and the CPT theorem. 

To obtain more concrete 
information from QFT we need to consider specific models.
For this we have a large degree of arbitrariness.
We are free to choose the dimension of spacetime, the spin
of the particles, the corresponding gauge group,
of arbitrary rank, the number of different `matter' fields of spin
smaller than $1$ and their corresponding representation under
the gauge group. Finally we are free to choose the couplings
among those fields, renormalizable or not, including the
 potential for the scalar fields, gauge and Yukawa couplings, etc.

Given that degree of degeneracy, we need to use some 
experimental input in order to choose an appropriate QFT that
could describe our world at least up to a given scale.
Such a particular example is the current standard model of 
particle physics based on the gauge symmetry $SU(3)\otimes SU(2)
\otimes U(1),$ with three families of quarks and leptons.
This model describes fundamental physics up to a scale of
$10^2$ GeV where $SU(2)\otimes U(1)$ is broken to electromagnetic
$U(1)$. We want to emphasize that this is only one in
an infinite number of QFTs and there is no reason other than experimental success
to select this model.

Nevertheless, it is widely believed that the standard model is only
an effective QFT that has to be generalized to a more 
fundamental theory. The main reasons for this belief are:

\begin{description}
\item{(i)} Gravitation is not described at the quantum level.
This is probably the most important problem of theoretical high energy
physics.

\item{(ii)} The gauge hierarchy problem which, 
roughly speaking, refers to the fact  that the $10^2$ GeV scale
of symmmetry breaking is not stable under radiative corrections.
In the presence of gravity it reduces
to the question of why this scale is so small compared with 
the Planck scale of $10^{19}$ Gev.

\item{(iii)} There are also the `why' problems (Why do we live in 
4D?, Why is the gauge group  $SU(3)\otimes SU(2)
\otimes U(1)$?, Why  are there three families?, Why do the couplings
and masses of the matter fields take the particular values
found by experiments?  Why is the cosmological constant essentially zero?
 etc.).
\end{description}

Extensions of the standard model in terms of QFTs are many, and
they partially 
 address  some of the problems mentioned above. For instance,
supersymmetric field theories represent the best candidates
to solve the hierarchy problem because the existence of a fermion-boson
symmetry can  stabilize the $10^2$ GeV scale\cite{super}. There are also 
grand unified theories (GUTs), which unify the gauge couplings
by imposing a simple gauge symmetry such
as $SU(5)$, broken  to the standard
model  at higher energies, and also the Kaluza-Klein theories
where it is assumed that the world is actually higher-dimensional
and the origin of gauge symmetries may be the extra small
dimensions of spacetime. All of these extensions of the standard
model are just different choices of QFTs and do not 
address the main problem of theoretical high energy physics,
namely the quantization of gravity.

String theory is the only candidate for a fundamental theory
of nature, encompassing all the known particles and interactions.
In particular, it is the candidate theory for a consistent treatment of gravity
in the quantum domain.
String theory is not just another extension of the standard 
model in terms of a QFT, it  is a generalization of 
the QFT itself. Roughly speaking the idea in string theory is
to replace the point-like particles of QFT by one-dimensional extended 
objects, strings, which could be open or closed.
Consistency requirements are very strict in string theory, selecting
only five supersymmetric theories in 10D, namely: type I
with gauge symmetry $SO(32)$, closed type II (A and B depending
on orientation properties) and closed heterotic theories
with gauge symmetries $SO(32)$ or $E_8\otimes E_8$.
Similar to QFT, string theory has very few general `predictions':
There is an infinite tower of massive states corresponding
to the oscillation modes of the string after quantization,
the masses are quantized in terms of a fundamental scale
which is identified with the Planck scale. 
On each of the five different theories there is always
a massless particle of spin 2, the graviton
$G_{MN}$, therefore strings {\it imply} the
existence of gravity. Type I and heterotic have also
massless particles of spin $1$, $A_M$,  implying the existence of gauge
symmetries and are therefore  candidates
to be `the fundamental theory of nature'. The spectrum  also 
includes a singlet massless scalar,
the dilaton $\Phi$ and antisymmetric tensors of different ranks
depending on the string.

The effective theories describing the massless particles
(the masses of observable particles are expected to arise
from the ordinary Higgs effect at lower energies)
are ordinary QFTs. We can see that string theories then 
lead to very restricted QFTs, with well defined 
symmetries and matter content. On the other hand each string theory
has many (thousands or billions of) vacua. This allows us to
construct string models in any number of dimensions less than $10$,
including quasi-realistic models in 4D, which are very similar to the
standard model, a very encouraging result. 
However, it also increases the level of arbitrariness, restricting
  the predictive power of the
theory
although, of course, the arbitrariness is still much less than 
in pure QFT's.

The natural way to construct 4D string models is to start with a 
10D theory and use the Kaluza-Klein idea of compactifying 
six dimensions in small spaces. Most of the variety of 
different vacua comes from
the freedom to choose among these 6D spaces, then
4D string models are sometimes called superstring compactifications
(SSC).
We can also substitute the 6D space by some 2D conformal field
theory (CFT) with specific properties as we will see in the next 
section.

4D string models are up to now the only candidates for
a fundamental theory of nature.  Their consistency   also 
requires supersymmetry, which is a welcome property
in view of the favourite solution to  the hierarchy problem. Strings also
improve the `why' problem by changing many of them into a single
dynamical question: Why do we live on this particular 
string vacuum or SSC?  
 This situation is again much better than QFTs, but it is not
completely satisfactory. We are left with at least  two unsolved problems:
\begin{description}
\item{(i)} How is supersymmetry broken, in order to recover the 
(non-supersymmetric) standard
model at low energies?
\item{(ii)} How do we lift the vacuum degeneracy and select one
single SSC describing our low energy world?
\end{description}
Fortunately, string theory is not yet completely understood, and
this is not the final status of the theory. String theory is
only understood at the perturbative level and non-perturbative questions, 
such as the
tunnelling effect or possible comparison of different vacua, 
cannot be approached at the moment. A non-perturbative 
formulation of the theory is expected to give an answer to 
the why question above, it is also expected to provide the mechanism
for the breakdown of supersymmetry at the electroweak scale,
 hopefully  maintaining almost vanishing  cosmological constant.

In these lectures I will review the field of string phenomenology.
I will briefly mention the different attempts to construct a quasi-realistic
4D string model, including the obstacles that have been found so 
far to obtain a realistic model. Next I will
discuss what is known about 4D effective actions
including the tree-level Yukawa couplings as well 
as the one-loop corrections to the gauge couplings and
nonrenormalization theorems. In chapter 4, I will present
some vacuum independent general results of 4D strings. This is
 the closest we can get to real predictions
of 4D string theories so far. Finally I will discuss briefly the
problem of supersymmetry breaking and the possible use of duality 
symmetries in approaching this and other string problems.

Since this subject is so vast, I have to restrict to a very superficial 
discussion; general introductions to string theory and CFT can be found
in \cite{gsw}. There are also two collections of some of the
more relevant papers on the subject of 4D strings. Ref.
\cite{bert1}, includes the string model building techniques
known before 1989, whereas ref.\cite{dine}, contains some
of the earlier phenomenological discussions of string
theory. A number of review articles touching some of the topics
of these lectures are provided in \cite{review}
  
\vspace{4.0 mm}
{\large\bf\noindent String Model Building }
\normalsize
\vglue 0.4cm

We mentioned in the introduction that there are only five 
consistent superstring theories in 10D. To build 
string models is the same as explicitly constructing 
the string vacua of each of these theories. By this we
mean solutions of the corresponding background field equations
of the different massless modes of the string.

Since there is no second quantized formulation of string theory, we need
to use first quantization. In this case the basic quantity is the
2D worldsheet action, which for the bosonic string is:

\beq
S=\frac{1}{\alpha'}\int d\sigma d\tau\left\{ \left(G_{MN}(X)+B_{MN}(X)\right)\partial^\mu X^M\,\partial_\mu X^N
+\alpha'\Phi(X)\, ^{(2)}R\right\}.
\eeq

Let us describe the different quantities entering into this action.
First the integral is over the 2D surface swept by the movement of 
the string.
This surface is parametrized by $\sigma,\tau$.
The inverse string tension $\alpha'$ is the only 
(constant) free parameter of the theory.
$X^M(\sigma,\tau), M=1, \cdots , D$ play two different roles: they are scalar fields
in the  2D theory,  but they are  coordinates of the target space where the
string propagates, which for critical string theories
(the subject of this paper) has
dimension $D=26$. Similarly $G_{MN}(X), B_{MN}(X), \Phi(X)$ are 
couplings of the 2D theory but since they are functions of $X$ they
are fields in  target space. $G_{MN}$ is a symmetric tensor
which is identiified with the metric; $B_{MN}$ is an antisymmetric tensor
field which in 4D target space will give rise to an axion field; and
$\Phi$ is a scalar field, the dilaton. Since it appears only 
multiplying the 2D curvature $^{(2)}R$ whose integral is the
topological invariant that counts the genus (number of holes) of the
corresponding 2D surface,  the vev of the dilaton is identified with the
string coupling. These fields are always present in any closed string.
% For heterotic strings we should also add
% gauge fields $A_M(X)$ which for consistency reasons are restricted
%to the gauge groups $SO(32)$ or $E_8\otimes E_8$.

A fundamental symmetry of the above action is conformal invariance  which 
includes scalings of the 2D metric as well as  
2D reparametrization invariance. Imposing this symmetry at the
2D quantum level is  similar to imposing that the coupling constants do not 
run in standard field theory. This  then defines
a 2D conformal field theory (CFT) and the constraints on the
2D couplings are the field equations for the target space fields
$G_{MN}, B_{MN}, \Phi, A_M$. Not surprisingly the
constraints give rise to Einstein's equations, Yang-Mills equations
and equations of motion for $B_{MN}$ and $\Phi$. To leading order in
$\alpha'$ these are the equations derived from the following
target spacetime effective action:

\beq
S=\int d^DX\sqrt{G}e^{-\Phi}\left\{R-\frac{1}{12}\nabla_M B_{NP}
\nabla^M B^{NP}+\nabla_M\Phi\nabla^M\Phi -\frac{D-26}{3}\right\}.
\eeq

Since heterotic strings are supersymmetric, we have to add the corresponding 
fermionic partners of those fields. Solutions of these equations
are then what we call string vacua and thus we can claim that there is
a correspondence between string vacua and certain CFTs in 2D.

The simplest solution is of course 26D flat spacetime with constant values
of all the fields. For this case we have a 2D free theory, which 
can be easily quantized by solving the wave equation $\partial^\mu\partial_\mu X^M=0$,
the fields $X^M$ can be written as:
\beq
X^M(\sigma,\tau)=X_R^M(\tau-\sigma)+ X_L^M(\tau+\sigma)
\eeq
as usual, $X_R^M$ and $X_L^M$ represent
right- and left-moving modes of the string respectively, with the mode expansion
\beqa
X_R^M(\tau-\sigma)=x_R^M+p_R^M(\tau-\sigma)+\frac{i}{2}\sum_{n\neq 0}   {\frac{1}{n}\alpha_n^M\, e^{-2in(\tau-\sigma)}}\nonumber\\
X_L^M(\tau+\sigma)=x_L^M+p_L^M(\tau+\sigma)+\frac{i}{2}\sum_{n\neq 0}   \frac{1}{n}\tilde\alpha_n^M\, e^{-2in(\tau+\sigma)}
\label{modos}
\eeqa

%The definition of the heterotic string is that the left 
%moving sector is purely bosonic and then has 
 Since this is a free theory, quantization assigns
canonical commutation relations to the Fourier 
coefficients $\alpha_m^N,\tilde\alpha_m^N$,  like oscillators of the
harmonic oscillator. The Hamiltonian then gives rise to the mass
formula:
\beq
%m_R^2=N_R-1,\qquad m_L^2= N_L-1.
M^2=N_R+N_L-2.
\label{masa}
\eeq
Where $ N_{R,L}$ refer to the harmonic oscillator occupation numbers for
left and right movers and the level matching condition requires
$N_L=N_R$ for consistency. Note that the `vacuum' state ($N_L=N_R=0$) is a tachyon
 and the next state requires one left-moving and 
one right-moving oscillator ($N_L=N_R=1$), since both oscillators carry a target space index, 
the state corresponds to an arbitrary two-index tensor $\alpha_{-1}^M
\tilde\alpha_{-1}^N |0\rangle$ of which the symmetric part is the
metric $G_{MN}$, the antisymmetric part is $B_{MN}$ and the trace is
the dilaton $\Phi$. That we can see are massless and are always present.

The instability due to the tachyon can be easily cured by supersymmetrizing the theory. In that case the tachyon state is projected out. The most popular
supersymmetric string theory is the heterotic string. In this theory, only the
right moving modes have a fermionic partner and consistency requires that
they live in a 10D space rather than the 26D space of the bosonic string.
The left moving modes however are purely bosonic, but the 26D space of these
modes is such that the extra 16 coordinates are toroidally compactified, giving rise to
extra massless states, which in this case are vector-like, as we will see next, and correspond
to the gauge fields of $SO(32)$ or $E_8\otimes E_8$.

\vspace{4.0 mm}
{\it\noindent Toroidal Compactifications}
%{\large\bf\noindent String Model Building }
\normalsize
\vglue 0.4cm

In order to construct string models in less than 10D as well as to
understand the heterotic string construction, we need to 
consider the simplest compactifications which correspond to the extra
dimensions being circles and their higher dimensional generalization.

Let us first see   the case of a circle. This means that the 10D space
is represented by flat 9D spacetime times a circle $S^1$. We know that a
circle is just the real line identifying all the numbers differing by $2\pi R$,
where $R$ is the radius of the circle.
So the only difference with the flat space discussed above are the boundary conditions. The solution of the wave equations are now as in (\ref{modos}).
But now  $p_R=m/2R-nR$ and $p_L=m/2R+nR$, $m$ is an integer
reflecting the fact that the momentum in the compact 
direction has to be quantized in order to get single-valued 
wave function. The integer $n$ however refers to the fact that
the string can wind around several times in the compact dimension and is thus
named the `winding number'. The mass formula is then: 
\beq
M^2=N_R+N_L-2+\frac{m^2}{4R^2}+n^2R^2, \qquad N_R-N_L=mn.
\eeq
This shows several interesting facts.
First, for $n=0$ and varying $m$, we obtain an infinite
tower of massive states with masses $\sim 1/R$; these are the
standard `momentum states' of Kaluza-Klein compactifications
in field theory. In particular the massless states with
$n=m=0$ and one oscillator in the compact direction
are vector fields in the extra dimensions giving rise
to a $U(1)_L\otimes U(1)_R$ Kaluza-Klein gauge symmetry.  The states with $n\neq 0$ are 
the winding states and are purely stringy; they represent 
string states winding around the circle, they have
mass $\sim R$.
Second,  there are special values of $m$ and $n$ which can give rise
to extra massless states. In particular for $m=n=\pm 1$ we can see that 
at the special radius $R^2=1/2$ in units of $\alpha'$, there are massless states with a single
oscillator $N_R=1, N_L=0$ corresponding to massless vectors which in this
case generate $SU(2)_R\times SU(2)_L$. This means that the special point
in the `moduli space' of the circle $R^2=1/2$ is a point of enhanced symmetry.
The original $U(1)_R\times U(1)_L$ Kaluza-Klein symmetry of compactification on a circle gets enhanced to $SU(2)_R\times SU(2)_L$. This is a very stringy effect because it depends crucially on the existence of winding modes
($n\neq 0$).
The third interesting fact about this compactification is that the spectrum 
is invariant under the following `duality' transformations \cite{gpr}:
\beq
R\leftrightarrow\frac{1}{2R}\qquad m\leftrightarrow n.
\label{dualp}
\eeq
This is also a  stringy property. It exchanges small
with large distances but at the same time it 
exchanges momentum (Kaluza-Klein) states with 
winding states. This symmetry can be shown to hold not
only for the spectrum but also for the interactions and
therefore it is an exact symmetry of string perturbation theory. %which 
%is now called $T$-duality.

\begin{figure}                                 
\begin{center}                                 
\leavevmode                                 
\epsfxsize=4in
\epsfysize=2.5in                                 
\epsffile{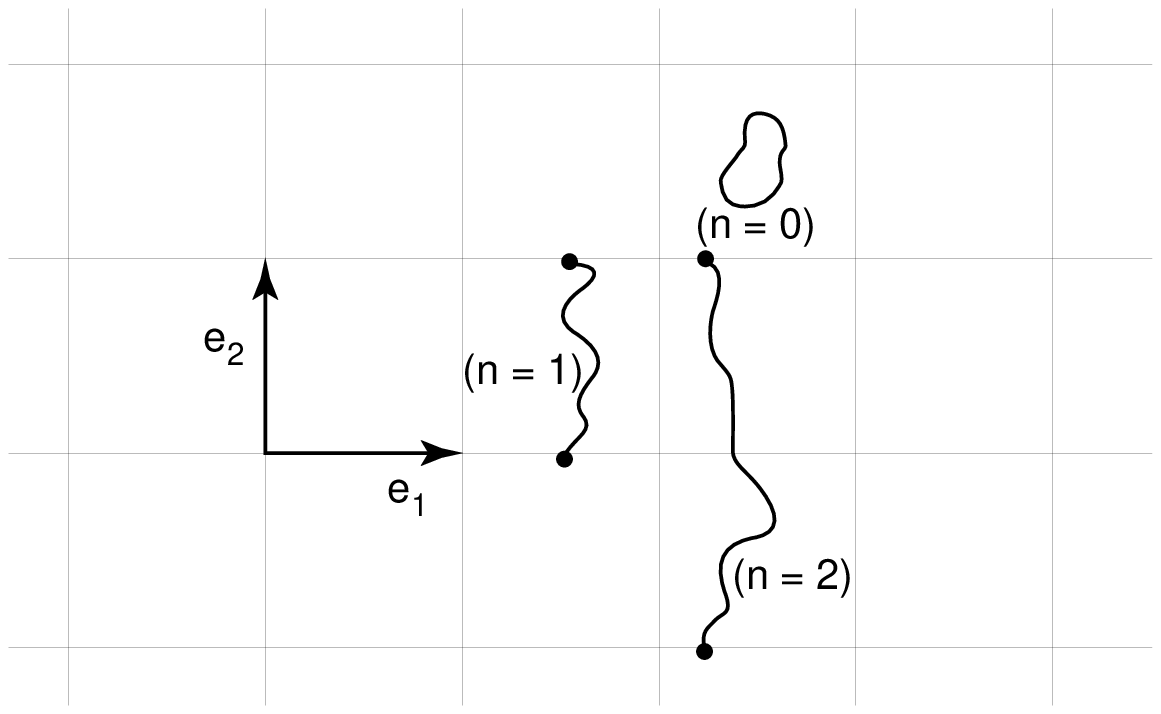}                             
\end{center}                                 
\caption{A 2D torus $T^2$ defined by the identification of points on $\IR^2$ by elements of the lattice defined by ${\bf e_1}$ and ${\bf e_2}$. We display
examples of a closed string on $\IR^2$ which is also
closed on $T^2$ ($n=0$), also a string closed on $T^2$ but not
on $\IR^2$, winding around the torus once($n=1$) and twice ($n=2$). \label{fig:mexfigtor}}
\end{figure}

Let us now   extend the compactification to two dimensions, {ie}
the 26D spacetime is the product of flat 24D spacetime and a
2D generalization of a circle, the torus $T^2$. %This compactification can
%also be analyzed since we have still free field thoeries in 2D.
Again the only difference with flat space is the boundary
conditions. The two compact dimensions are identified by
vectors of a 2D lattice, defining the torus $T^2$.
Out of the three independent components of the compactified
metric $G_{11}, G_{22}, G_{12}$ and the single component of
$B_{MN}$ namely $B_{12}$ we can build two complex `moduli' fields:
\beqa
U\equiv \frac{G_{12}}{G_{22}}+i\, \frac{\sqrt G}{G_{22}}\nonumber\\
T\equiv B_{12}+i\, \sqrt G.
\eeqa
$U$ is the standard modular parameter of any geometrical
 2D torus and it is usually identified as the `complex 
structure' modulus. $T$  is the `K\"ahler structure' modulus
(since $T^2$ is a complex K\"ahler space) and its imaginary part 
measures the overall size of the torus, since $\sqrt G$ is the
determinant of the 2D metric. It  plays the same role as $R$
did for the 1D circle. In terms of $T$ and $U$ we can write the left- and
right-moving momenta as:
\beqa
p_L^2=\frac{1}{2U_2T_2}\|(n_1-n_2\, U)-T\, (m_2+m_1\, U)\|^2\nonumber\\
p_R^2=\frac{1}{2U_2T_2}\|(n_1-n_2\, U)-T^*(m_2+m_1\, U)\|^2
\label{momenta}
\eeqa
The mass formula, depending on $p_L^2+p_R^2$, again shows that there are enhanced symmetry points for
special values of $T$ and $U$. It also shows the following symmetries:
\beq
U\rightarrow\frac{a\, U+b}{c\, U+d}\qquad T\rightarrow\frac{a\, T+b}{c\, T+d}
\qquad T\leftrightarrow U.
\eeq
Where $a,b,c,d$ are integers satisfying $ad-bc=1$.
The first transformation is the standard 
$SL(2,\IZ)_U$ `modular' symmetry of 2D
tori and is independent of string theory; it is purely geometric.
The second transformation is a stringy $SL(2,\IZ)_T$ named 
$T$-duality and it is a generalization of (\ref{dualp})
for the 2D case. Again this is a symmetry as long as we also transform
 momenta $m_1,m_2$ with winding $n_1,n_2$.
 The third symmetry exchanges the
complex structure $U$ with the K\"ahler structure $T$ and
it is called `mirror symmetry'.
If $U$ and $T$ each parametrize a complex plane
$SL(2,\IR)/O(2)$, the duality symmetry implies that they can only live
in the fundamental domain defined by all the points of the product of complex spaces $SL(2,\IR)/O(2)\otimes SL(2,\IR)/O(2)\cong O(2,2,\IR)/(O(2)\times O(2))$ identified under the 
duality group $SL(2,\IZ)_U\times SL(2,\IZ)_T
=O(2,2,\IZ)$.

This is the situation that gets generalized to higher dimensions.
In general, compactification on a $d$-dimensional torus has the
moduli space ${\cal M}=O(d,d,\IR)/O(d)\times O(d)$ with 
points identified under the duality group $O(d,d,\IZ)$.
For the heterotic string with 16 extra left moving coordinates
${\cal M}=O(d+16,d,\IR)/O(d+16)\times O(d)$ with
a similar modification to the duality group.
The left- and right- moving momenta $p_L, p_R$ live on an
even, selfdual lattice of signature $(22,6)$, which
is usually called the Narain lattice $\Lambda_{22,6}$\cite{narain}.
This generalizes the $\Lambda_{2,2}$ lattice defined by the
integers $m_1,m_2; n_1,n_2$ of eq. (\ref{momenta}).

We can easily verify in this case that the dimension of
${\cal M}$ is $d(d+16)$ corresponding to the number of independent
components of $G_{mn}, B_{mn}, A_{m}^I$ with 
$m,n=1\cdots d; I=1,\cdots 16$. For $d=6$ we have a 
4D string model with a moduli space of dimension $132$. To
this we have to add the dilaton field $\Phi$ which, together
with the spacetime components of the antisymmetric
tensor $B_{\mu\nu}$, can be combined into a new modular
parameter:
\beq
S\equiv a+i\, e^{\Phi}.
\eeq
Here the axion field $a$ is defined as $\nabla_\mu a=
\epsilon_{\mu\nu\rho\sigma}\nabla^{\nu}B^{\rho\sigma}$.
 $S$ parametrizes again a coset $SL(2,\IR)/O(2)$. It is then natural to
believe there is also a duality symmetry for the field $S$ 
of the type $SL(2,\IZ)$; by analogy with the situation for $T$ and $U$.
Such a symmetry was proposed in ref.\cite{filq2}\ and it
has received a lot of attention recently. If true it 
may have far reaching
consequences since (similar to equation (7)) it relates
strong to weak string coupling.

%\ifig\fone{ 
%Potential etc. etc..
%}{\epsfxsize=5.6cm \epsfysize=5.1cm \epsfbox{mexfig1.ps}}

\vspace{4.0 mm}
{\it\noindent Orbifold Compactifications}
\vglue 0.4cm

 We have then succeeded in constructing 4D superstring models
from toroidal compactifications and understand the full
class of these models given by the moduli space ${\cal M}$.
Unfortunately, all of these models have $N=4$ supersymmetry
and therefore they are not interesting for phenomenology,
because they are not chiral. To obtain a chiral model we should
construct models with at most $N=1$ supersymmetry.
If we still want to use the benefits of free 2D theories,
we should construct models from flat space and modify only the
boundary conditions. We have already considered identifications
by shift symmetries of a lattice defining the tori. We still have the
option to also use rotations and consider `twisted'
boundary conditions \cite{orbifolds}. As an example let us start with the
torus $T^2$ discussed before. % it is defined
%by the identification 
% $X^i\rightarrow X^i+k_1 e_1^i+k_2 e_2^i$
%where $e_\alpha^i$ are the two basis vectors defining the square
%lattice.
 If we make the identification
$X^i\rightarrow -X^i$ we are constructing  the orbifold
$O^2\equiv T^2/\IZ_2$, shown in figure 2, where the $\IZ_2$ twist is rotation by 
$\pi$. This space {\it is not} a manifold because it is singular at the points left fixed  by the rotation  $\{(0,0),(0,1/2), (1/2,0), (1/2,1/2)\}$.
Notice that, for instance,  the point $(1/2,1/2)$ is fixed because it is transformed to
$(-1/2,-1/2)$ which is identical to the original point after
a lattice shift. In general, the discrete group of rotations defining
the orbifold is  called the point group
${\cal P}$, whereas the
nonabelian group including the rotations and also the 
 translations  of the lattice $\Lambda$, is  the space group ${\cal S}$.
So usually a torus is defined as $T^d\equiv \IR^d/\Lambda$ and
an orbifold $O^d\equiv T^d/{\cal P}\equiv \IR^d/{\cal S}$.

\begin{figure}                                 
\begin{center}                                 
\leavevmode                                 
\epsfxsize=5in
\epsfysize=1.5in                                 
\epsffile{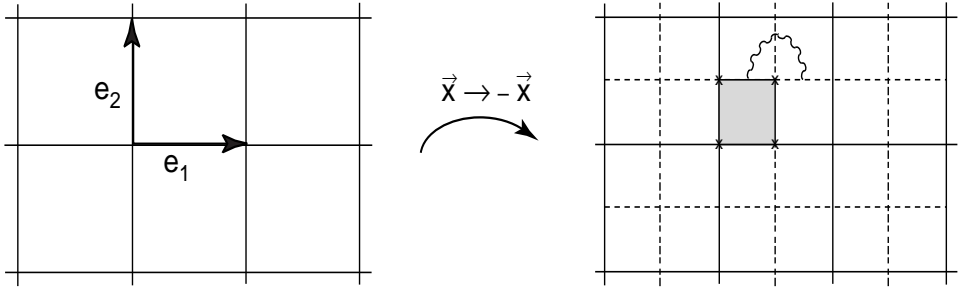}                             
\end{center}                                 
\caption{Starting from the two-torus $T^2$, we generate the orbifold
$O^2\equiv T^2/\IZ^2$, `the ravioli', by the identification $\vec x\leftrightarrow
-\vec x$. $O^2$ is singular at the four `fixed points' shown. Besides the momentum and winding states of the torus, the orbifold spectrum also has `twisted' states, corresponding to strings closed in $O^2$
but not on $T^2$. The twisted states are attached to fixed points, we display one example. \label{fig:orb}}
\end{figure}

We can easily construct 4D strings from orbifold compactifications
in which the 10D spacetime of the heterotic string  is the product of 4D flat spacetime and a
six-dimensional orbifold $O^6$. The heterotic string is particularly interesting
because we can extend the action of the point group to the 16D 
lattice of the gauge group by embedding the action of the orbifold twist
in the gauge degrees of freedom defined by the $E_8\times E_8$ lattice,
say. This can easily be done in two ways:

\begin{description}
\item{(i)} Perform a homomorphism of the point group action in 
the gauge lattice by {\it shifting} the lattice vectors by a vector
$V=ML$ where $M$ is the order of the point group and $L$ is
any lattice vector in 16D.

\item{(ii)} Perform the homomorphism by twisting also the 
gauge lattice by an order $M$ rotation belonging to the
Weyl group of the corresponding gauge group.

\end{description}

These embeddings on the gauge degrees of freedom allow us to
break the gauge group, reduce the number of supersymmetries and 
generate $N=1$ chiral models in 4D as desired.
The reason for this is the following:
using the embedding by a shift $V$, we start with the
spectrum of the toroidal compactification and have to
{\it project out} all the states that are not invariant by the
orbifold twist. For the gauge group, only the elements
satisfying $P\cdot V\in \IZ$ remain, where $P\in E_8\otimes E_8$, breaking the gauge group to a subgroup of the same
rank. The four gravitinos of the $N=4$  toroidal compactification 
also transform and depending on the orbifold twist 
they are reduced to only one or two invariant states, indicating
that there is only $N=2$ or $N=1$ supersymmetry.
Actually there are only four twists $\IZ_M$ leading to 
$N=2$ (for $M=2,3,4,6$) and some twenty $\IZ_M$ or $\IZ_M\times\IZ_N$
twists leading to $N=1$ supersymmetry
\cite{zetan}, which are the phenomenologically 
interesting ones. For each of these twists we can have several
($\sim 10$) different embeddings on the gauge degrees of freedom.
One of these embeddings is called the standard embedding because
it acts identically in the gauge degrees of freedom as in 
the 6D space, this embedding also describes
compactifications of the type II strings and is distinguished
because in the 2D worldsheet, the corresponding model has two
supersymmetries on the left-movers  and two supersymmetries on the
right-movers, the corresponding models
are called $(2,2)$ models. All other embeddings do not have
supersymmetry in the left moving side and are called $(0,2)$ models.

On top of all these embeddings we can also add Wilson lines
\cite{wilson}\
$A_m, m=1,\cdots,6$, by embedding the shifts
of the lattice defining the 6D compactified torus, on the
gauge degrees of freedom in terms of further shifts
of the 16D gauge lattice, which will
further break the gauge group. This increases the number of
possible consistent models by a large amount, which   we
can only estimate  between millions and billions
because
 some may turn out to be actually equivalent.
The Wilson lines can also be interpreted as the full embedding
of the space group ${\cal S}$ in the gauge degrees
of freedom. In this case, using the two alternative embeddings mentioned
above will
give a completely different result because in the first case, both $V$
and the Wilson lines $A_i$ will act as shifts and so the embedding is
abelian, whereas in the second option we will have both shifts
and twists so the embedding is non-abelian. This possibility 
allows for two important properties: the Wilson lines are continuous
rather than quantized and the rank of the gauge group can be reduced.
In the absence of Wilson lines both embeddings are
equivalent.
Both classes of embeddings can be obtained by starting with the
Narain lattice of toroidal compactifications $\Lambda_{22,6}$ and 
twist it in a consistent manner. This already takes into account the discrete
and continuous Wilson lines (which were already present,
parametrizing $\Lambda_{22,6}$) and also allows for the possibility of
performing left-right asymmetric twists, the so-called asymmetric
orbifolds
\cite{asymmetric}. This extra degree of freedom increases the number of
possible models.

We can see now how a vast amount of heterotic string 
orbifold models can be generated. There are many (billions?) classes
of models; different classes differentiated by the choice of
original 6D toroidal lattice, the orbifold point group, the embeddings
$V$ and the discrete Wilson lines. But each of these {\em discrete}
choices
allows for a  variation of different {\em continuous} parameters 
such as  the
moduli fields (like $S,T,U$), the continuous Wilson lines
(that correspond to charged untwisted sector moduli fields) and 
there are also
 charged twisted-sector moduli fields\cite{mirjamus}.
All the continuous parameters can be seen as flat potentials
for fields in the effective field theoretical effective action.

Only a few of these classes of models 
includes quasi-realistic models. As an example
\cite{standard}, the
model based on the $\IZ_3$ orbifold with embedding 
$V$ and nonvanishing Wilson lines $A_m$ given by:
\beqa
V=\frac{1}{3}\left(1,1,1,1,2,0,0,0\right)\times \left(2,0,\cdots,0\right)\nonumber\\
A_1=A_2=\frac{1}{3}\left(0,0,\cdots,2\right)\times \left(0,1,1,0\cdots,0\right)\nonumber\\
A_3=A_4=\frac{1}{3}\left(1,1,1,2,1,0,1,1\right)\times\left(1,1,\cdots,0\right)
\eeqa
breaks $E_8\times E_8$ to $SU(3)\times SU(2)\times U(1)\times
U(1)^7\times SO(10)$ with three families of quarks and leptons.
The extra $U(1)^7$ symmetry can be broken by the standard 
Higgs mechanism which, in string theory requires the existence of flat directions among some charged matter fields; these can be analyzed in the models at hand because there are general `selection rules'
forbidding couplings not invariant under the action of the point and
space groups. The $SO(10)$ remains as a hidden sector
(in the sense that it only has gravitational
strength couplings with the observable $SU(3)\times SU(2)\times U(1)$ sector).
This is an example of a quasi realistic model. The structure of Yukawa
couplings can be analyzed leading to very realistic properties
and problems such as very fast proton decay can be avoided.
However, there are extra doublets in the model that 
give rise to unrealistic values of Weinberg's angle.
This may in principle be solved by contemplating the existence of intermediate
scales, but  at this point the model stops being
stringy. It  also has the drawback that without knowing details
about supersymmetry breaking many of the low energy parameters
can not be determined. There are variations of this model that allow    for an extra $U(1)$ symmetry at low energies, implying
a relatively light $Z'$ particle.
There are several models in the literature with similar properties as this one, showing that it is possible to get
models very close to the standard model of particle physics. But there
is not a single model that could be considered realistic.
In particular there is no model yet with just the spectrum of the
supersymmetric standard model.

\vspace{4.0 mm}
{\it\noindent Calabi-Yau compactifications }
\vglue 0.4cm

We saw that the orbifolds obtained from twisting the 6D tori
can give rise to chiral $N=1$ models in 4D. Orbifolds 
are singular objects but they can be smoothed out by 
blowing-up  the singularities at the fixed points.
The resulting smooth manifold is a so-called Calabi-Yau
manifold \cite{calabi}. Mathematically, these are 6D complex manifolds with $SU(3)$
holonomy or equivalently vanishing first Chern class.
They were actually the first standard Kaluza-Klein compactification
considered in string theory, leading to chiral 4D models
and generically gauge group $E_6\times E_8$, with $E_8$ a 
hidden gauge group.

The drawback of compactifications on Calabi Yau manifolds
is that they are highly nontrivial spaces and 
we cannot describe the strings on such manifolds, contrary to what
we did in the case of free theories such as tori and orbifolds.
In particular we can not compute explicitly the couplings
in the effective theory, except for the simplest renormalizable
Yukawa
couplings.

On the other hand, Calabi-Yau manifolds have been understood
much better during the past few years and have lead to some
beautiful and impressive results. In a way they are more general than orbiifolds because an orbifold is only a particular
singular limit of a Calabi-Yau manifold. Also there are
other constructions of these manifolds which are
not related to orbifolds. They can be defined as
hypersurfaces in complex (weighted) projective spaces 
$\IP^4_{(k_0,k_1,k_2,k_3,k_4)}$ where $k_i$'s are the weights
of the corresponding coordinates for which there is the
identification $z_i\cong \lambda^{k_i} z_i$. The hypersurface is
defined as the vanishing locus of a polynomial of the 
corresponding coordinates. For instance
the surface defined as:
\beq
P\equiv z_1^{12}+z_2^{12}+z_3^6+z_4^6+z_5^2=0
\eeq
 defines a Calabi-Yau manifold with weights:$(1,1,2,2,6)$.
The relation $\sum_i k_i=d$ where $d$ is the degree of the polynomial
ensures that the surface is a Calabi-Yau manifold. The manifold
is guaranteed to be smooth if both the polynomial $P$ and its derivatives
do not vanish simultaneously.
 Larger classes of manifolds can be
constructed by considering intersections of hypersurfaces in higher
dimensional projective spaces, the so-called complete intersection Calabi-Yau
manifolds (CICY). Actually, it is known in the mathematical literature, that
 {\it all} Calabi-Yau spaces can be defined as
(intersection of) hypersurfaces in weighted projective spaces.  Large numbers of these manifolds have been
classified, although the full classification is not complete.
The mathematical estimate is that there are of the order
of  ten  thousand Calabi-Yau manifolds. The corresponding string models
are of the type $(2,2)$.
Some of the highlights of Calabi-Yau compactifications are
the following:

\begin{description}

\item{(i)} There are classes of moduli fields, generalizing
the fields $T$ and $U$ of the two-torus mentioned before.
The number of these fields is given by topological numbers
known as Hodge numbers $h_{i,j}, i+j\leq 3$. 
They correspond to the number of 
complex harmonic forms that can be defined in the manifold
with $i$ holomorphic indices and $j$ antiholomorphic
indices. Then the number of complex structure fields (U)
is given by $h_{2,1}$ and the number of K\"ahler structure
fields ($T$) is given by $h_{1,1}$.
Many of the $h_{2,1}$ forms correspond to the coefficients of 
different monomials which can be added to the defining polynomial 
that still give rise to the same space, other forms correspond
to the blowing up of possible singularities. Many of the $h_{2,1}$ forms
correspond to polynomial deformations of the defining surface,
but others are related with the reparation of singularities.
For
Calabi-Yau manifolds we have $h_{1,1}\geq 1$, therefore there is
always a special K\"ahler class deformation which can be thought 
as the overall size of the corresponding manifold, it
is usually called, the K\"ahler form. All the other Hodge numbers
of Calabi-Yau manifolds are fixed ($h_{3,0}=h_{0,3}=1, h_{1,0}=h_{0,1}=0,h_{0,2}=h_{2,0}=0$)

\item{(ii)} The gauge group in 4D is $E_6\times E_8$.
The matter fields transform as ${27}$'s or ${\ov  27}$'s
of $E_6$. The number of each is also given by the Hodge numbers
$h_{1,1}, h_{1,2}$ and the number of generations is then
 topological:$N_g\equiv h_{1,1}- h_{1,2}=\chi/2$ where
$\chi$ is the Euler number of the manifold. This is one of the
most appealing properties of these compactifications since they
imply that topology determines the number of quarks and leptons.

\item{(iii)}Mirror symmetry \cite{mirror}. Similar to the 2D toroidal compactifications,
it has been found that there is a mirror symmetry in Calabi-Yau 
spaces that exchanges the moduli fields $T_m$ and $U_a$
$m=1,\cdots h_{1,1}$, $a=1,\cdots, h_{2,1}$. This means that for every 
Calabi-Yau manifold ${\cal M}$, there exists another  
manifold ${\cal W}$ which has the complex and K\"ahler structure
fields exchanged, {\it ie} $(h_{1,1}, h_{2,1})\leftrightarrow
(h_{2,1},h_{1,1})$ and opposite Euler number $\chi$.
The mirror symmetry of the two-torus described previously,
is only   a special example
on which the manifold  is its own mirror.
Mirror symmetry is not only a nontrivial contribution
of string theory to modern mathematics, but it has very interesting
applications for computing effective Lagrangians as we will see in 
the next chapter. it also relates the geometrical
modular symmetries associated to the $U_a$ fields of the manifold ${\cal M}$
to generalized, stringy, $T$-duality symmetries for the mirror
${\cal W}$ and viceversa (see for instance \cite{cdgp}\ ).

\item{(iv)} Even though these string models are not completely understood in 
terms of 2D CFTs, special points in the moduli space of a given
Calabi-Yau are CFTs, such as the orbifold compactifications mentioned
before. In fact some people believe that there is a one to one
correspondence 
between  string models with $(2,2)$ supersymmetry in the worldsheet and
Calabi-Yau manifolds. There is also a description of CFTs in terms of
effective Landau-Ginzburg Lagrangians which are intimately related 
to Calabi-Yau compactifications as two phases of the same 2D theory
\cite{landau}.
In particular the potential of that Landau-Ginzburg theory is
determined by the polynomial $P$ defining the Calabi-Yau hypersurface.

\item{(v)} There are also few models with three generations that,
after the $E_6$ symmetry breaking, could lead to quasi-realistic
4D strings. One of these models was analyzed in some detail
\cite{graham}.
Usually, $E_6$ models also lead to the existence of extra $Z'$ particles
of different kinds. They have been thoroughly studied because of
the potential experimental importance of detecting an
extra massive gauge boson (for a recent discussion see ref. \cite{mirlang}\ ).

\item{(vi)} Although most of the Calabi-Yau models studied so far
correspond to standard
embedding in the gauge degrees of freedom
($(2,2)$ models), there is also 
the possibility of constructing $(0,2)$ models by performing 
different embeddings, similar to the orbifold case.
This increases substantially the number of string models
of this construction \cite{zerotwo}.
\end{description}
 
\vspace{4.0 mm}
{\it\noindent Other Constructions}
\vglue 0.4cm

During the past few years several other constructions
of chiral 4D strings in four dimensions, have been found
in terms of explicit CFT's. We described before how the use of
free field CFT's lead us naturally to orbifold
compactifications. We can also use the property of these
2D field theries for which there is an equivalence 
between fermions and bosons. Since the bosonic fields in 
2D are the coordinates in target space, by fermionizing
them we lose the geometrical interpretation, but it is
a consistent string model as long as we keep the 4D spacetime
coordinates as bosons. If we fermionize all the extra coordinates
and choose nontrivial boundary conditions on the fermions we can
get nontrivial 4D strings \cite{fermions}, which do not have to have
a geometrical interpretation in terms of compactifications!. Many models
have been studied using this approach which, in many cases, are
equivalent to orbifold compactifications at some particular value
of the radius. Some quasi-realistic models have been
studied with much detail using this approach. This
includes models with three families and standard model gauge group
$SU(3)\times SU(2)\times U(1)$ as well as a version of
$SU(5)\times U(1)$ known as flipped $SU(5)$
\cite{flip}. In some  cases
again, the models reproduce many  of the nice features of the
standard model. Nevertheless, as in the case of orbifolds, there
is not a totally realistic model yet.

A related approach uses bosonization in the opposite direction
{\it ie} it bosonizes all of the fermions of the (supersymmetric)
right moving sector (including the ghost system needed for consistent quantization of the 2D theory)
\cite{bosons}. This is the so called covariant lattice
approach which in some way generalizes the Narain lattice of toroidal
compactifications. Again many of these models are equivalent to
orbifolds at a particular radius. In particular some of the three
generation orbifold models mentioned before have been explicitly reproduced
using this approach \cite{bos3}.

A probably more general construction goes into the name
of Gepner-Kazama-Suzuki models \cite{gepner}. They depart from free field
CFTs and construct more general CFTs by using cosets
$G/H$ to describe the CFT of the `internal' dimensions.
 This construction includes (products of) statistical mechanics models
such as the Ising and Potts models and their supersymmetric
generalizations. One of the salient features of these constructions
is that for the models with $(2,2)$ supersymmetry in the
worldsheet, it can be shown explicitly that they do correspond
to particular points of Calabi-Yau compactifications, despite of their
original non-geometric construction. This was found by using their
realization in terms of Landau-Ginzburg 2D effective field theories.
Generalization of these constructions to $(0,2)$ supersymmetric
models have also been achieved and a large class of models exist
\cite{cerodos},
including  some close to the standard model.

We have then seen that there are several formalisms for
constructing chiral 4D strings. Many models can be built
using different approaches. Each formalism has
advantages and disadvantages in terms of the level 
of generality and for performing explicit calculations for
the low energy effective theory. 

\vspace{4.0 mm}
{\large\bf\noindent Effective Actions in 4D }
\vglue 0.4cm

We have seen that consistent 4D chiral string models lead
to $N=1$ supersymmetry. For phenomenological purposes,
we are interested in finding the effective action for the
light degrees of freedom, that means we want to integrate out
all the heavy degrees of freedom at the Planck scale $M_p$
and compute the effective couplings among the light 
states (massless at the Planck scale).
This will be
a standard field theoretical action with $N=1$ supersymmetry.
The on-shell massless spectrum of these models
have the graviton-gravitino multiplet $(G_{\mu\nu},\Psi_\mu)$,
the gauge-gaugino multiplets $(A_\mu^\alpha,\lambda^\alpha)$
and the matter and moduli fields which fit
into $N=1$ chiral multiplets of the form $(z,\chi)$
except for the dilaton field which together with the 
antisymmetric tensor $B_{\mu\nu}$ belong to a linear 
$N=1$ multiplet $L=(\Phi,B_{\mu\nu},\rho)$. 
The most general couplings of supergravity to one linear
multiplet and several gauge and chiral multiplets is
{\it not} yet known, although some progress towards its construction has been done recently \cite{linear,linearus}. Nevertheless, as we mentioned in the
previous chapter, this field can be dualized to construct the
chiral multiplet $(S,\chi_S)$ with $S=a+i\, e^\Phi$ and
$\nabla_\mu a\equiv \epsilon_{\mu\nu\rho\sigma}\nabla^\nu
B^{\rho\sigma}$. After performing this duality transformation we are
lead with a $N=1$ supergravity theory coupled only  to gauge  
and  chiral multiplets.
The most general such action  was constructed more than a decade ago
\cite{cremmer},
and therefore it is more convenient in this sense to work with the dual dilaton $S$ rather
than the stringy one $L$. Although, the partial knowledge
about the lagrangian in terms of $L$ is enough to understand
most of the results we will mention next \cite{linearus}, we will
only mention the approach with the field $S$ which is 
the most commonly used.
The general Lagrangian coupling $N=1$ supergravity 
to gauge and chiral multiplets
 depends on three arbitrary functions of the chiral multiplets:

\begin{description}
\item{(1)}The K\"ahler potential $K(z,\ov z)$ which is a {\it real}
function. It determines the kinetic terms of the chiral 
fields 
\beq
{\cal L}_{kin}=K_{z\ov z}\partial_\mu z\partial^\mu \ov z
\eeq
with $K_{z\ov z}\equiv \partial^2 K/\partial z\partial\ov z$.
% with subindices
%reflecting derivatives with respect to the corresponding variable.
$K$ is called K\"ahler potential because the manifold
of the scalar fields $z$ is K\"ahler, with metric $K_{z\ov z}$.
 
\item{(2)}The superpotential $W(z)$ which is a {\it holomorphic}
function of the chiral multiplets (it does not depend on
$\ov z$)\footnote{Actually, W is  a section of a line bundle \cite{bw}.}. $W$ determines the Yukawa couplings as well
as the $F$-term part of the scalar potential $V_F$ (known as $F$-term 
because it originates after eliminating auxiliary fields associated to the chiral multplets which are usually
called $F$):
\beq
V_F(z,\ov z)=e^{K/M_p^2}\left\{ D_zW\, K^{-1}_{z\ov z}\ov{D_zW}-3\frac{|W|^2}{M_p^2}\right\},
\eeq
with $D_zW\equiv W_z+WK_z/M_p^2$. Here and in what follows, the internal
indices labelling different chiral multiplets $z_i$ are not
explicitly written.

\item{(3)}The gauge kinetic function $f_{ab}(z)$ which 
is also {\it holomorphic}. It determines the gauge kinetic terms
\beq
{\cal L}_{gauge}= \re f_{ab} F_{\mu\nu}^a\, F^{\mu\nu\, b}
+\im f_{ab} F_{\mu\nu}^a\, \tilde F^{\mu\nu\, b}
\eeq
it also contributes to gaugino masses and the gauge
part of the scalar potential (coming from eliminating 
the auxiliary fields $D$ of the gauge multiplets and referred to as
$D$-terms).
\beqa
V_D &=&\left(\re f^{-1}\right)_{ab}\left(K_z, T^a \, z\right)\, \left(K_{\ov z}, T^b\, \ov z\right)\nonumber\\
V &=& V_F+V_D
\eeqa
 
\end{description}

The problem posed in this section is:
given a 4D string model, calculate the functions
$K,W,f$. In order to do that let us
separate the fields into the moduli $U,T,$
the dilaton $S$,
\footnote{From this section on, we will depart from the
conventions of the previous sections about the definitions of $S,T,U$,
the only change is to make $M\rightarrow iM$ for $M=S,T,U$ then
the axionic component is now the imaginary component of the complex field and 
the compactification size is the real component of $T$. The reason for this change is to have consistency with the standard supergravity conventions.
In particular, the $SL(2,\IZ)_T$ duality is now the transfoamtion $T\rightarrow (aT-ib)/(icT+d)$.}
 and the matter fields  charged under the gauge group $Q^I$.
Most of the structure of the couplings depends
on the model. But there are some couplings which are model independent.
To extract them, the best procedure is to use all the
symmetries at hand. For this let us remark that 4D strings
are controlled by two perturbation expansions. One is the
expansion in the sigma-model (2D worldsheet) which is governed by the
expectation value of a modulus field $T$ (the size of the extra dimensions).
Whereas proper string perturbation theory is governed by the dilaton field $S$.

\vspace{4.0 mm}
{\it\noindent Tree-level Couplings }
\vglue 0.4cm

Let us consider first the couplings generated at string tree-level
and also tree-level in the sigma-model expansion \cite{wittentrunc,bfq}.
Besides the 4D Poincar\'e symmetry, supersymmetry and
gauge symmetries which determine the Cremmer {\em et al} Lagrangian,
we also can use the `axionic' symmetry:
$B_{MN}\rightarrow B_{MN}+{\rm closed form}$
This is a symmetry which for the 4D fields would
imply that $S,T,U$ can be shifted by an  arbitrary imaginary constant.
There are also two scaling properties of the 4D Lagrangian
$S\rightarrow \lambda S$, $G_{\mu\nu}\rightarrow
\lambda G_{\mu\nu}$ for which the Lagrangian
scales as ${\cal L}\rightarrow \lambda {\cal L}$.
Also, given a scale $\Lambda$, define $\tau=\kappa\Lambda^4$
where $\kappa$ gives Newton's constant in 10D. The transformations
$S\rightarrow \tau^{-1/2} S$, $T\rightarrow \tau^{1/2} T$
with similar transformations for the other fields, imply that the
Lagrangian should scale as ${\cal L}(\kappa)\rightarrow \tau^{-1/2}{\cal L}
(\Lambda^{-4})$. These scaling properties are not symmetries of the
Lagrangian but of the classical field equations and so 
they can be used to restrict the form of the 
{\it tree-level} effective action only.

Using these symmetries we can extract the full dependence of the
effective action on the dilaton field $S$, which is the most 
generic field in all compactifications. We conclude that at tree-level
in both expansions \cite{bfq}:
\beqa
K(S,T,U,Q^I)&=&-\log(S+S^*)+\hat K(T,U,Q)\nonumber\\
W(S,T,U,Q^I)&=&y_{IJK}Q^{I}Q^{J}Q^{K}\nonumber\\
f_{ab}(S,T,U,Q^I )&=&S\, \delta_{ab}
\label{treetree}
\eeqa
With $\hat K$ still undetermined.
This is however a very crude approximation. What we really want
is to know these functions at tree-level in the string expansion but
{\it exact} in the sigma model expansion. This should be 
achievable because many of the 4D models are exact
2D CFTs as we saw in the previous chapter.
We can still extract very useful information from
equation (\ref{treetree}).
As we said above, the axionic symmetries  imply
that to {\it all orders} in sigma-model
expansion the superpotential does not depend on 
$T,U$ and it is just a cubic function of the matter fields
$Q^I$. This is important for several reasons:
First, we know the field $T$ comes from the internal components of
the metric and  controls the loop expansion of the worldsheet
action. If $W$ does not depend on $T$ it means that it cannot get 
any corrections in sigma model perturbation theory!\cite{zerotwo}. Therefore 
the only $T,U$ dependence of the (exact) tree-level
superpotential is due to nonperturbative effects in the worldsheet, in particular 
all nonrenormalizable couplings in the superpotential are exponentially suppressed ($\sim e^{-T}$)\cite{dsww}.
A way to see that there are nonperturbative worldsheet corrections to the
string tree-level superpotential is to realize that 
the axionic symmetry shifting $T$ by an imaginary constant,
 is broken by nonperturbative worldsheet
effects to $T\rightarrow T+i\,n,\,  n\in\IZ$.
This is nothing but
one of the $SL(2,\IZ)_{T,U}$ transformation for toroidal orbifold
compactifications ($a=b=d=1, c=0$ in eq. (10)). Therefore the only conditions these symmetries impose on $W$
is that it should transform as a modular form of a given weight
($W\rightarrow (cT+d)^{-3}\, W$ for the simplest toroidal orbifolds with
$T$ the overall size of the compactification space)\cite{flst}.
In fact, explicit calculations for specific orbifold models show that
\beq
W_{tree}(T,Q^I)=\chi_{IJK}(T)\, Q^{I}Q^{J}Q^{K}+\cdots
\eeq
with $\chi(T)$ a particular modular form of $SL(2,\IZ)$ or any other duality group
 and the ellipsis
represent higher powers of $Q$, exponentially suppressed.
The identification of $\chi(T)$ with modular forms was a highly nontrivial 
check of the explicit orbifold calculations which were preformed in refs.
\cite{hv}\
without any relation (nor knowledge) of the
underlying duality symmetry $SL(2,\IZ)$. This kind of symmetry
puts also strong constraints to the higher order,
nonrenormalizable, corrections to $W$, since each matter field $Q$ transforms in a particular way under that symmetry
($Q\rightarrow(cT+d)^n\, Q$ with $n$ the modular weight of $Q$).
There are also other discrete symmetries, as those defined by the point
group ${\cal P}$ and space group ${\cal S}$ of an orbifold which have
to be respected by the superpotential $W$. These `selection rules' are
very important to find vanishing couplings and uncover flat directions
which can be used to break the original gauge symmetries and
construct quasi-realistic models.

%However for $T$ and $U$ this is only valid perturbatively in the sigma
%model expansion, the shift symmetry gets broken to  a discrete symmetry
%when the full CFT is taken into account. 
Second, and more important, the superpotential above does not depend
on $S$ which is the string loop-counting parameter, and therefore
$W_{tree}$ does not get renormalized in string perturbation theory!\cite{dsnr}.
This
means that we only need to compute $W$ at the tree level and it
will not be changed by radiative corrections. This is the string version of the standard non-renormalization theorems of supersymmetric theories.
Also for $Q=0$ the superpotential vanishes, independent of the values of
$S,T,U$ ($W(S,T,U,Q=0)=0$)! There are not self
couplings among the `moduli' fields and therefore they represent
flat directions in field space
(see for instance \cite{lance}\ ). Notice that due to the non-renormalization theorems, this result is exact in string perturbation theory!.
The only possibilty we have to lift this vacuum degeneracy is by 
nonperturbative string  effects.

The quantity we have less information on, even at tree-level, is the
Kahler potential $\hat K(T,U,Q)$. It has been computed only for several
simple cases. For instance in the simplest possible Calabi Yau
compactification ($h_{1,1}=1, h_{2,1}=0$) a consistent
truncation from the 10D action gives \cite{wittentrunc}:
\beq
K=-\log(S+S^*)-3\log(T+\ov T+Q \ov Q)
\eeq
Curiously enough, the second term appeared in the so-called
`no-scale models' studied before string theory \cite{noscale}.
This form holds also for the untwisted fields of orbifold compactifications,
but the dependence on the twisted fields is not known.
It also gives the appropriate result in the large
radius limit of Calabi-Yau compactifications although
it gets non-perturbative worldsheet corrections relevant at
small radii.
In order to find the exact tree-level K\"ahler potential, 
the best that has been done so far is to write the
K\"ahler potential as an expansion in the matter fields \cite{dkl}:
\beqa
K &=&- \log(S+\ov S)+K^M(T,\ov T, U,\ov U)+K^Q(T,\ov T, U,\ov U)Q\ov Q\nonumber\\
&+ &Z(T,\ov T, U,\ov U)(QQ+\ov Q\ov Q)+{\cal O}(Q^3),
\eeqa
and compute the moduli dependent quantities $K^M,K^Q,K^E$.
This has been done explicitly for some $(2,2)$ orbifold
compactifications.
For instance for factorized orbifolds,
that is orbifolds of
a 6D torus which is the product of three 2D tori $T^2$,
the dependence on the corresponding moduli 
fields is given by
\beqa
K^M &=& -\sum_a\log(T_a+\ov T_a)-\sum_m\log(U_m+\ov U_m),\nonumber\\
K^Q &=& \prod_{a,m}\left(T_a+\ov T_a\right)^{n_m^I}\, \left(U_m+\ov U_m\right)^{n_a^I},
\label{potk}
\eeqa
and $Z(T,\ov T, U,\ov U)=0$. Giving rise to the Kahler potential: 
\beqa
K &=& -\log(S+\ov S)-\sum_a\log(T_a+\ov T_a)-\sum_m\log(U_m+\ov U_m)\nonumber\\
& &+\sum_I \left|Q_I\right|^2\prod_{a,m}\left(T_a+\ov T_a\right)^{n_m^I}\, \left(U_m+\ov U_m\right)^{n_a^I}
\label{potkc}
\eeqa
Where the fractional numbers $n_m^I, n_a^I$ are the
`modular weights' of the fields $Q_I$ with respect to the 
duality symmetries related to the moduli $T_a$ or $U_m$.
For instance, under $T$ duality, the fields $Q^I$ transform as:
\beq
Q^I\rightarrow (ic_m T_m+d_m)^{n_m^I}\, Q^I
\eeq
 
Furthermore, for $(2,2)$ (Calabi-Yau) models, there is 
a very interesting observation
\cite{seiberg}. Since these compactifications are also
compactifications of type II strings, and in that case
the corresponding 4D theory has $N=2$ supersymmetry, the moduli
dependent part of the K\"ahler potential, has to have the
same dependence as for the $N=2$ models which are much more restrictive.
 This underlying $N=2$ structure has been very fruitful to extract 
information on $(2,2)$ models and comes with the name of
`special geometry'.
This restricts the function $K_M$ which gives the metric in the moduli
space. First, the moduli space
of the $h_{1,1}$ forms $T_a$ and the $h_{2,1}$ forms $U_a$
factorizes and so:
\beq
K^M(T,\ov T, U,\ov U)=K^T(T,\ov T)+K^U(U,\ov U)
\eeq
For which the eq. (\ref{potk}) is a particular case.
Second,
 in $N=2$ supergravity, the full Lagrangian is 
completely determined by a single holomorphic function, the prepotential. The K\"ahler potential for the $T_a$ fields is a function of the
prepotential $F(T_a)$, given by \cite{neq2}:
\beq
K^T(T,\ov T)=-\log\left(F+\ov F-\frac{1}{2}\left(F_T-\ov{F_T}\right)
\left(T-\ov T\right)\right)
\eeq
Where on the right hand side, the subindices mean 
differentiation. A similar expression holds for $K_U$ in terms of a second
prepotential $G(U_m)$.
Furthermore, the moduli dependence of the cubic terms
in the superpotential
\beq
W_{tree}=\frac{1}{3}W_{abc}(T_a) Q^aQ^bQ^c+\frac{1}{3}W_{mnp}(U_m)\hat Q^m\hat Q^n\hat Q^p+\cdots ,
\eeq
is also given by the functions $F(T_a)$ and $G(U_m)$ since the
Yukawa couplings are given by
\footnote{The physical Yukawa couplings depend also on  the K\"ahler potential,
see for instance \cite{at}.} 
\beqa
W_{abc}(T_a)&=&\partial_a\partial_b\partial_c
F(T_a) \nonumber \\
W_{mnp}(U_m)&=&\partial_m\partial_n\partial_p G(U_m).
\eeqa
Since $F$ and $G$ are holomorphic, they may get similar constraints as
 the superpotential above.
In particular, since
 $T$ counts sigma model loops, then $G(U)$ is not renormalized
and so
the $U$ dependent part 
of the K\"ahler potential $K_U(U,\ov U))$ is given exactly by the
tree-level result!. Similarly,  the Yukawa couplings
$W_{mnp}(U_m)$ are exact at tree-level.
Here is where the mirror symmetry explained in the previous 
section plays an important role. Since by mirror symmetry
we understand that the compactification on the original manifold
${\cal M}$ and its mirror ${\cal W}$ represent the
same CFT, and so the same string model, in the version
with the manifold ${\cal W}$, the roles of $T$ and $U$
are interchanged, therefore, computing the $U$ dependent part 
of the K\"ahler potential in ${\cal W}$
(which is exact at tree level) gives the
$T$ dependent part of the K\"ahler potential in ${\cal M}$!
This fact has been used to compute explicitly the
moduli dependent part of the K\"ahler potential in some
examples. This  overcomes the Calabi-Yau 
problem of not knowing the exact CFT behind the
compactification, at least for these couplings.

\vglue 0.4cm
\vspace{4.0 mm}
{\it\noindent Loop Corrections}
\vglue 0.4cm

We have seen there is  good understanding of some
of the tree-level couplings of 4D string models.
Also non-renormalization theorems guarantee that
the superpotential computed at tree-level is exact 
at all orders in string perturbation theory.
This powerful result depends crucially on the fact that
the superpotential is a holomorphic function of the fields,
so if by the Peccei-Quinn symmetry cannot depend on the imaginary
part of the dilaton field $S$, then it cannot depend neither
on the real part of $S$. This fact cannot be used for
the K\"ahler potential, which in general will be corrected
order by order in string perturbation theory.
This is then the least known part of any string theory
effective action.
On the other hand the gauge kinetic function $f$ is
also holomorhic and we know it exactly at the tree-level
($f=S$). Since this function determines the
gauge coupling itself, it is very interesting to
consider the loop corrections to $f$.

During the past several years, explicit one-loop
corrections to $f$ have been computed, especially for
some orbifold models.
First it was shown that string loop diagrams reproduce the standard
running of the gauge couplings in field theory, 
as expected. More interesting though, was to find the
finite corrections given by threshold effects which 
include heavy string modes running in the loop.
These corrections will be functions of the moduli fields, such as the
geometric moduli $T,U$ but also other mooduli such as 
continuous Wilson lines of orbifold models.

For factorized orbifold models, 
 the explicit dependence of the one-loop corrections
 on the
$T_i, U_i$ moduli fields takes the form \cite{dkl}:
\beq
f_a=k_a\, S-\sum_i\frac{\alpha^i_a}{4\pi^2}\log \eta(iM^i)+{\rm constant}
\eeq
where $k_a$ are the Kac-Moody levels of the correpsonding gauge groups, The coefficients $\alpha^i_a $ are group theoretical quantities depending on the Casimir $T(\Phi)$ of the representation of the matter fields $Q$ and on the Casimir of the adjoint $T(G_a)$ as well as on the modular weights of the
fields $Q^I$. %($Q^I\rightarrow \, \prod_i(i c_i M^i+d_i)^{-n^i_I}Q^I$.
\beq
\alpha^i_a=\sum_IT_a(Q^I)(1-2n^i_I)-T(G_a).
\eeq
$M^i$ refers to the set of both moduli $U,T$ for each of the three tori.
and $\eta$ is the Dedekind function:
\beq
\eta(iT)\equiv e^{-\pi iT/12}\prod_{n=1}^\infty \left(1-e^{2\pi i n T}\right).
\eeq
Notice that although $\eta(M)$ transforms in 
a simple way under $SL(2,\IZ)$ transformations, the function
$f$ is not invariant under $T,U$ duality.
This is as well because the quantitity we need 
to be invariant is not $f$ but the physical gauge coupling which
depends not only on $f$  but also on the K\"ahler potential at tree-level.
%We can explicitly verify that it is invariant.
Notice the singularity structure of $f$.
Using arguments about singularities and invariance of
the full gauge coupling, it is possible to extract the
expression of the function $f$ in more complicated cases such as
Calabi-Yau compactifications, for which the
string computation is not possible and the duality grup is not
as simple as $SL(2,\IZ)$ \cite{kl}. Also. extra modifications
to equation (29) have been computed recently, including its
dependence on continuous Wilson lines \cite{dkl}.

The calculation of the string one-loop corrections to the gauge couplings is better compared to the Lagrangian for the dilaton in a linear multplet $L$.
After performing the duality transformation it was found that those are not only corrections to the function $f$ but also to the K\"ahler potential
$K$ of the form
(for factorized orbifolds)\cite{dfkz}:
\beq
K_{1-loop}=-\log\left(S+\ov S+\frac{1}{4\pi^2}\sum_{i=1}^3\delta_i^{GS}\log(T_i+\ov T_i)\right)-
\sum_{i=1}^3\log(T_i+\ov T_i)
\eeq
Where the coefficients $\delta_i^{GS}$ are related to a Green-Schwarz counterterm, cancelling duality
anomalies, and have been explicitly computed for different
orbifolds. This expression for $K$ has been verified by other explicit 
one-loop calculations,  to the first order in a 
$(S+\ov S)^{-1}$ expansion \cite{agn}.

The knowledge of loop corrections to $f$ is not only important for studying 
questions of gauge coupling unification and supersymmetry breaking
by condensation of hidden sector gauge fermions.
It is also important because there is also a non-renormalization theorem
for $f$ staying that {\em there are no further
corrections to f beyond one-loop}
\cite{in}. This is again as in standard
supersymmetric theories \cite{russians}. The only thing to keep in mind is that 
for $f$ it is important to state clearly that
we are working with the `Wilson' effective action rather than the 1PI
effective action.
In this case the gauge kinetic function is holomorphic
and does not get renormalized beyond one loop \cite{russians}.
On the other hand, the 1PI gauge coupling is {\it not} holomorphic
and does get corrections from higher loops but, since it gives
the physical coupling, it is invariant under duality
symmetries.

\vspace{4.0 mm}
{\large\bf\noindent Model Independent Results }
\vglue 0.4cm

Let us in this section recapitulate what we can say about string models 
which is independent of the model.
This is the closest we can get to string predictions and
help us in approaching general questions, differentiating the
generic issues from those of a particular model.
Since the full nonperturbative formulation of string theory is not yet
available, we have to content ourselves mostly with 
predictions of string perturbation theory, assuming that
the corresponding string model is given buy a CFT.

\begin{description}
\item{(i)}
First, as it was mentioned in the introduction, 4D string models
{\it predict} the existence of gravity and gauge interactions.
This is a point that cannot be overemphasized since it is the
first theory that makes those fundamental predictions for interactions
we experience in the every day life.
The dimension of spacetime is dynamical and $D\leq 10$ raising the
hope that eventually we could explain if a 4D spacetime
is in some way  special, although at present, the signature of spacetime is fixed in string theory as it is in QFT. Also
the rank of the gauge group is bounded $r\leq 22$
\footnote{Both of the last two statements have been recently
modified at the non-perturbative level, 
by the studies of strong-weak coupling duality
symmetries (for a recent review see \cite{joe}\ ). In particular there is some evidence for the
existence of an 11D theory from which all the different strings
could emerge. There is also evidence for the appearance
of nonperturbative gauge groups that can raise the rank beyond $r=22$
\cite{witteninst}.} 

\item{(ii)}There are other fields which 
survive at low energies: charged matter fields, candidates to be basic blocks of matter but also the dilaton field $S$ and other moduli
$T,U$. We have to mention  that, although as yet there
is no 4D model without moduli fields, there is not
a general theorem implying their existence. In that sense the dilaton is the most generic modulus field, with a flat potential
in perturbation theory.

\item{(iii)}There is only one arbitrary parameter 
$\alpha'$ fixed to be close to the Planck scale
$M_p$. All other parameters
of the effective action are determined by expectation values of fields
such as the dilaton and the moduli. In particular the gauge coupling is
given at tree level by the vev of $S$.

\item{(iv)} The existence of spacetime supersymmetry is needed for consistency,
although $N=1$ is selected for phenomenological reasons.
There is a general requirement for a CFT to lead to $N=1$ spacetime supersymmetry: It has to have $(0,2)$ supersymmetry in the wordlsheet
(2D) (plus  a quantization condition on the
charges of the $U(1)$ group mixing the two supersymmetries)
\cite{bdfm}.
Furthermore, it is not possible to break supersymmetry 
perturbatively nor in a continuous matter by the smooth variation of 
a parameter
\cite{dsmac,bd}. \footnote{This last claim depends on some CFT assumptions 
\cite{bd}\ or that the breaking of supersymetry comes from an $F$-term
\cite{dsmac}.
There could be 
counterexamples evading the assumptions.}

\item{(v)} There are no global internal symmetries in 4D string
models
\cite{bd}, besides the already mentioned Peccei-Quinn symmetry of
the $S$ field and some accidental global symmetries
(like baryon and lepton numbers in the standard model).
This is a very strong result derived by showing that if there is
an internal symmetry, the properties of CFT's imply that there should
be a vector field in the spectrum with the properties of the gauge field of that
symmetry. This is consistent with similar claims about the
nonexistence of global symmetries in gravity, due for instance to 
wormhole effects. This puts very strong constraints to string models
compared with standard field theory models.

\item{(vi)}There are generically some {\it discrete} symmetries
in string models. Some infinite dimensional
such as $T$-duality and others finite dimensional as those inherited from the point 
group of orbifold constructions,  which are seen as
discrete gauge symmetries. These can in principle be useful for
model building, hierarchy of masses etc. There are however some couplings
that vanish in string theory and {\it cannot} be explained in terms of symmetries of the effective 4D theory, these are called `string miracles'
since from the 4D point of view they seem to break the criterium for naturalness \cite{dsmir}.
As we saw in the previous section, $T$-duality symmetries restrict 
very much the form of the effective action and quantities such as 
Yukawa couplings have to be modular forms of a given duality group.
These symmetries are valid to all orders in string perturbation theory
and are thought to be also preserved by nonperturbative effects.
Matter fields $Q^I$ are assigned special quantum numbers,
the modular weights $n$, according to their transformation properties under the duality group. For a $SL(2,\IZ)^m$ group for instance we have:
\beq
Q^I\rightarrow (ic_l\, T_l+d_l)^{n^l_I}\, Q^I,\quad l=1,\cdots ,m.
\eeq
Since fermions transform nontrivially under these symmetries, there
may be `duality anomalies' which have to be cancelled for consistency.
This imposes strong constraints on the possible spectrum of the
corresponding string model (which has to satisfy $\alpha^i_a/k_a=$
constant). For instance, using this, it can be shown that 
it is impossible to obtain the minimal supersymmetric standard
model spectrum  from any $\IZ_3$ or $\IZ_7$ orbifold models
\cite{il}.
Anomaly cancellation also implies that the dilaton $S$ has
to transform nontrivially under $T$-duality, for factorized orbifolds:
\beq
S\rightarrow S-\frac{k_a}{8\pi^2}\sum_i\delta^i_{GS}\, \log(c_iT_i+d_i).
\eeq

\item{(vii)}
There is unification without the need of a GUT. If 
the gauge group is a direct product of several groups we have
for the heterotic string
\cite{paul}:%we 
%assume the standard model group $SU(3)\times SU(2)\times U(1)$
%at low energies, the corresponding gauge couplings satisfy:
\beq
k_1\, g_1^2=k_2\, g_2^2=\cdots=\frac{8\pi}{\alpha'}G_{Newton}\equiv
g^2_{string}.
%\frac{k_1}{g_1^2}=\frac{k_2}{g_2^2}=\frac{k_3}{g_3^2}=\frac{1}{g_{grav}^2}
\eeq
Where $k_i$ are special stringy constants known as the Kac-Moody levels of the corresponding
gauge groups (for the standard model groups it is usually assumed that
$k_2=k_3=1, k_1=5/3$).
We can see there is a difference with standard GUTs in field theory
for which we compute the unification scale by finding the
point where the different string couplings meet.
In heterotic string theory, the unification scale is given in terms of the
string coupling $g_{string}$ and the 
Planck scale.\footnote{For type I strings 
the gravitational and gauge couplings are independent, so 
we have the freedom to adjust the unification scale with experiment
as in usual GUTs}
 More precisely:
$M_{string}\sim 5.27\times 10^{17} g_{string}$ Gev.
For $g_{string}\sim {\cal O}(1)$ this shows a discrepancy with the
`observed' value of the unification scale given by the
experiments $M_{GUT}\sim 2\times 10^{16}$ GeV. Also 
the Weinberg angle gives $\sin^2\theta_W=0.218$ differing from the
experimental value of $\sin^2\theta_W=0.233\pm0.0008$.
Therefore the string `predictions' are
very close to the experimental value, which is encouraging, but
differ by  several standard deviations from it. This is the 
string unification problem. The situation looks much better
for simple GUT's which have good agreement with experiment.
Several ideas have been proposed to cure this problem, including 
large values of threshold corrections, intermediate scales, extra
particles, changing the values of Kac-Moody levels etc
\cite{ilr}, with 
no compelling solution yet (for recent discussions of this
issue see for example \cite{stieberg} and the review of K. Dienes in ref. [5].).  
Recently, a strong coupling solution of this problem was proposed which leads to  a lower bound on Newton's constant, closed to the observed one \cite{wittenn}.

\item{(viii)} Most of the models have Kac-Moody level $k=1$ and therefore cannot have 
adjoint representation particles. In order to have GUTs from strings,
higher level models are then needed
\cite{dave}. These can be constructed from 
level-one models, for instance the breaking of $G\times G$ to
$G_{diagonal}$ leads to a level-two model.  Some three and four generation string- GUTs have been 
explicitly constructed. General CFT relations show that 
in order to implement the `missing partner'
and see-saw mechanisms in string GUTs we would need
Kac-Moody levels $\geq 5$. Also, if the breaking to the standard model
by an adjoint is a flat direction then states transforming as
$(8,1,0)+(1,3,0)+(1,1,0)$ under
$SU(3)\times SU(2)\times U(1)$ remain in the light spectrum
and could have low-energy implications.

\item{(ix)} There are usually fractionally charged particles in 4D string models \cite{fcww}.
In fact it can be shown that we cannot have simultaneously
$k_2=k_3=3k_1/5 =1$ in the standard model and only integer charged
particles, because if that is the case the standard model 
gauge group would be enhanced to a full level-one
$SU(5)$\cite{bert}.  This `problem' can be evaded in models where  the
fractionally charged particles are heavy string states, it has
also been proposed that those particles could confine at intermediate energies and 
be unobservable \cite{fc}.

\item{(x)}Nonrenormalization theorems of the previous sections
are very strong and are model independent. We know the superpotential is exact at
tree-level and the $f$ function is exact at one-loop. The K\"ahler
potential is however renormalized perturbatively.
These results imply that the lifting of flat directions and the breaking of
supersymmetry have to be achieved only at the string non-perturbative level (unless we do it by hand at tree-level \cite{ss}\ ).

\item{(xi)}
There are  `anomalous' $U(1)$ groups in most
of the models, but  there is also a counterterm in the action cancelling the anomaly and  generating a Fayet-Iliopoulos kind of term \cite{dsw}:
\beq
{\cal L_{FI}}=\frac{1}{S+\ov S}\left\vert\frac{Tr q^a}{48\pi^2}\frac{1}{(S+\ov S)^2}+\sum{q_I^a |Q_I| ^2}\right\vert^2,
\eeq
where $q_I^a$ are the anomalous charges of the scalar fields $Q_I$.
This term is responsable to break the would be anomalous group by fixing the 
value of a combination of the matter fields $Q_I$, breaking 
the would be anomalous $U(1)$ and usually other gauge groups, but not supersymmetry. A combination of the fields $Q_I$ and the dilaton $S$
still remains massless and plays the role of the new dilaton field.

\end{description}

There are also further model independent results which refer to nonperturbative
string effects and  will be discussed  next.

\vspace{4.0 mm}
{\large\bf\noindent Supersymmetry Breaking}
\vglue 0.4cm

In this chapter we will approach the problem of supersymmetry breaking in string theories in two complementary ways.
First we will discuss the favorite mechanism for low energy supersymmetry breaking, namely gaugino condensation, next we will 
try to extract general information about the effects of
supersymmetry breaking, independent of the particular mechanism
that breaks, we will then discuss the general form of soft breaking terms expected in string models, independent of the breaking mechanism.
 This is similar to the case of the standard model
where the Higgs sector can be treated as a black box but we can study the
low-energy theory below the symmetry breaking scale without
relying on the particular mechanism that broke the symmetry.
We will end up with general scenarios for the breaking 
of supersymmetry, discussing in particular the very generic problem, known as 
the `cosmological moduli problem'.

We have seen that in the efforts to extract a relation between string theory
and physics, there are two main problems, namely how
 the large vacuum degeneracy is lifted and how  
 supersymmetry is broken at low energies.
These problems, when present at string tree level,
cannot be solved at any order in string  perturbation 
theory. Therefore  the only
hope to solve them  is nonperturbative physics.
This has a good  and a bad side. The good side is that 
nonperturbative effects represent the most natural way to 
generate large hierarchies due to their exponential 
suppression, this is precisely what is needed to obtain the
Weinberg-Salam scale from the fundamental string or Planck scale.
The bad side is that despite many efforts, we do not yet have
 a nonperturbative formulation of string theory.
At the moment, the only concrete nonperturbative information
we can extract is from the purely {\it field theoretical}
 nonperturbative 
effects inside string theory (although a great amount of
progress has been made during the last year on nonperturbative string effects, 
\cite{joe}\ ). Probably the simplest and
certainly the most studied of those effects is gaugino
condensation in a hidden sector of the gauge group, since it
has the potential of breaking supersymmetry as well as lifting
some of the flat directions, as we will presently discuss.

\vspace{4.0 mm}
{\it\noindent {Gaugino Condensation}}
\vglue 0.4cm

The idea of breaking supersymmetry in a dynamical way was
first presented in refs.~\cite{witten}. In those articles a
general topological argument was developed in terms of the 
Witten index $Tr(-)^F$, showing that dynamical supersymmetry 
breaking {\it cannot} be achieved unless there is chiral matter or
we include supergravity effects for which the index
argument  does not apply.
This was subsequently verified by explicitly studying gaugino 
condensation in pure supersymmetric Yang-Mills, a vector-like
theory, for which gauginos condense but do not break
global supersymmetry \cite{vy} (for a review see \cite{amati}).
Breaking  global supersymmetry with chiral matter
was an open possibility in principle, but this approach ran into 
many problems when tried to be realized in practice.

The situation improved very much with the coupling to supergravity.
 The reason was that simple
gaugino condensation was argued to be sufficient to break 
supersymmetry once the coupling to gravity was included. This
 works in a hidden sector mechanism where gravity is the messenger
of supersymmetry breaking to the observable sector
\cite{peter}. However, it has recently been realized that the
proposal in \cite{peter} does not work
(see for instance \cite{mirev}\ which also includes an extended
discussion of the current status of gaugino condensation);  coupling with supergravity does not actually change the situation in global supersymmetry where gaugino condensation does not break supersymmetry. The missing ingredient was
the fact that gauge couplings were considered to be
constant rather than field dependent.
Gaugino condensation with field dependent gauge couplings was
anticipated in ref. \cite{fgn} and is realized in a very natural way 
in string thoery. As we have seen, the gauge coupling is a function of the
dilaton and moduli fields.
Furthermore,  string theory provides a natural realization of the
hidden sector models
 \cite{din,drsw}\ by having
 a hidden sector especially in the $E_8\times E_8$ versions.
 
To study the effects of gaugino condensation we should be
able to answer the following questions:
Do gauginos condense? If so, is supersymmetry broken by this 
effect? What is the effective theory below the scale
of condensation?
In order to answer these questions, several ideas have been put forward
\cite{vy,fgn,drsw,kp,mr, bdqq2}. 
The most convenient formalism is in terms of the so-called 2PI effective action
\cite{bdqq2}. In order to understand this formalism, it 
is convenient to think
about the case of spontaneous breaking of gauge symmetries.
In that case we minimize the effective potential for
a Higgs field, obtained
from the 1PI effective action and see if the
minimum breaks or not the corresponding gauge symmetry.
In our case, we are interested in the expectation value
of a composite field, namely $\lbarl$ or its
supersymmetric expression $\WW$. Therefore we need
the  two particle irreducible effective
action (2PI).% introduced in ref.~\cite{jt}. 
 
We start then with the generating functional in
the presence of an external current $J$ coupled to 
the operator that we want the expectation value of,
namely,
$\WW$. Let us consider the simplest case of a single
hidden sector gauge group in global supersymmetry.
(Coupling to supergravity presents no obstacles but 
make the discussion more cumbersome \cite{bdqq2,mirev}.)
\beq
e^{i{\cal W}[S,T,J]} \equiv \int DV \exp i\int d^4x\left\{
[(f(S,T) + J ) Tr\WW]_F +{\rm cc}\right\} 
\eeq
From this we have\footnote{The classical field $U$ defined here has no relation with the moduli fields $U$ of the previous sections.} 
\beq
\frac{\delta{\cal W}}{\delta J}=\Avg\WW\equiv U
 \eeq
and define the 2PI action as 
\beq
\Gamma[f(S,T), U]\equiv {\cal W}-\int d^4x\left( U J\right)
\eeq
To find the explicit form of $\Gamma$ we
use the fact that ${\cal W} $ depends on its two arguments
only through the combination $f+J$,
therefore, we can see that $\delta \Gamma/\delta f
=\delta \Gamma/\delta J= U$. Integrating this
equation implies that $\Gamma$ has to be {\em linear}
in $f(S,T)$\cite{bdqq2}:
%determines the dependence of $\Gamma$ in $f(S,T)$:
\beq
\Gamma[f(S,T), U]= U\, f(S,T)+\Xi ( U)
\eeq
where the arbitrary function $\Xi (U)$ can be determined using symmetry arguments as in \cite{vy,kl,bdqq2}.
We find:
\beq
W[f,U]=U[f(S,T)+\frac{c}{3}\log U+\xi]%\nonumber\\
%e^{-K/3}&=&e^{-K_p/3}-a\, \left(UU^*\right)^{1/3}
%\end{eqnarray}
\eeq
Here $\xi$ is an arbitrary constant. 
%and $K_p$ refers to the perturbative K\"ahler potential.
We can see that the superpotential corresponds to
the one found in \cite{vy}. 
Therefore this  2PI action is a reinterpretation of the
one in \cite{vy}. We have to stress that in our treatment
$U$ is only a {\it classical} field, {\it not}
 to be integrated out in any path integral. It also does not make sense to consider loop
corrections to its potential, this solves the question 
raised in \cite{mr} where loop corrections to the
$U$ potential could change the tree level results.
Furthermore, since $U$ is classical we can eliminate it by
just solving its field equations: $\d\Gamma/\d U=0$.
%(Since this implies $J=0$, it makes equations (11) and (9)
%reduce to (7).) 
These equations cannot
be solved explicitly but we find the solution in
an $1/\Lambda$ expansion, with $\Lambda$ the scale of condensation  \cite{bdqq2}. We find that the solution of these
equations reproduces at first order, 
 the Wilson action derived in \cite{drsw},
for which
we can just read the superpotential
(using $f(S,T)=S$)
to be:
\beq
%\begin{eqnarray}
W(S)=w e^{-3S/c} %\nonumber \\
%e^{-K/3}&=&e^{-K_p/3}-k\,e^{-(S+S^*)/c}.
\label{siuperp}
\eeq 
%\end{eqnarray}
where $w$ is an arbitrary constant.
%(to assure positive kinetic energy).
The superpotential is just the one found in \cite{drsw}. The
correction to the K\"ahler potential is not completely known
due mainly to the fact that the {\it perturbative} corrections to $K$ are
completely unknown
\cite{bdqq2}. Notice that these 
are corrections of  order $e^{-1/g^2}$ as expected.

By studying the effective potential for $U$ we recover the previously
known results. For one condensate and field independent
gauge couplings (no field $S$) the gauginos
condense ($U\neq 0$) but supersymmetry is unbroken.
For field dependendt gauge coupling, the minimum is for $U=0$
($S\rightarrow\infty$) so gauginos {\it do not} condense
(this is reflected in the runaway behaviour of the Wilsonian
action for $S$). %For several condensing groups we find $U\neq 0$
%and supersymmetry broken or not, depending on the case
%\cite{ccm}.

Alternatively, after eliminating $U$ from its field equations 
and using (\ref{siuperp}), we find the scalar
potential for the real parts of $S$ and $T$ ($S_R$ and $T_R$
respectively), namely $V(S_R,T_R)\sim \frac{1}{S_RT_R^3}\exp(-3S_R/4\pi b)$.
This potential has a runaway behaviour for both $S_R$ and $T_R$,
as expected.

 The $T$ dependence of the potential was completely changed after the consideration of target space or $T$ duality.  It was shown \cite{filq}, that imposing this symmetry
changes the structure of the scalar potential for the moduli fields
in such a way that it develops a minimum at $T\sim 1.2$ (in 
string units), whereas the potential blows-up at the decompactification
limit ($T_R\rightarrow\infty$), as desired (see figure 3)
\footnote{Notice that the scalar potential blows up at large radius which is {\it weak}
sigma-model  coupling. This is anti-intuitive, since we would have expected the potential to vanish at weak coupling. A way to understand this is to
realize that $1/g_4^2=R^6/g_{10}^2$ that means that for large $R$
and fixed $g_4$, the
original 10D string coupling becomes large, so the potential is blowing-up at strong string coupling
from the 10D point of view (we thank J. Polchinski and S.-J. Rey
for explaining this point).}.
The modifications due to imposing $T$ duality can be traced to the fact 
that the gauge couplings get moduli dependent threshold corrections from
loops of heavy string states \cite{dkl} as in eq. (29). This in turn generates a 
moduli dependence on the superpotential induced by gaugino condensation
of the form 
\beq
W(S,T)\sim\eta(iT)^{-6}\exp(-3S/8\pi b),
\eeq
 with 
$\eta(T)$ the Dedekind function \footnote{This formula is actually  more complicated
if the coefficients $\delta^{GS}\neq 0$ in which case $S$ also transform under T-duality as in eq. (34), see for instance B. de Carlos {\it et al} in 
\cite{krasnikov}}. 

\begin{figure}                                 
\begin{center}                                 
\leavevmode                                 
\epsfxsize=2.9in
\epsfysize=2.6in                                 
\epsffile{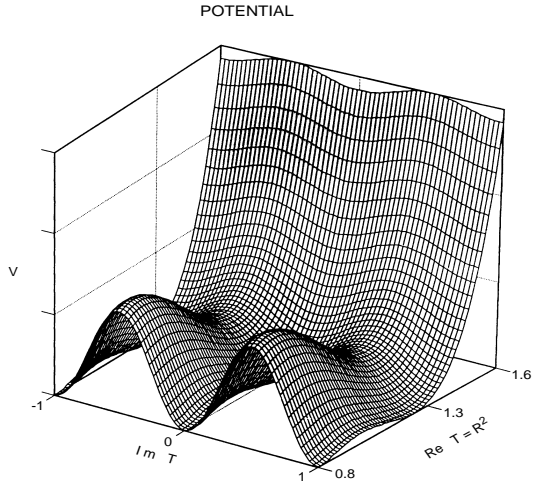}                             
\end{center}                                 
\caption{ The $T$-duality invariant potential from gaugino condensation. Notice that the potential blows-up at large compactification radius and
the minimum is at small radius ($R^2\sim 1.2$ in string units) as
desired. \label{fig:pot }}
\end{figure}

This mechanism however did not help in changing the runaway behaviour
of the potential in the direction of $S$.
There is a very generic problem emphasized mostly by Dine and Seiberg
\cite{dsprob} . It is known that because at large $S$ the string is weakly coupled, the potential has to vanish asymptotically (towards a free theory).
 Any other minimum has to
be at strong coupling for which the perturbation expansion does not work, unless there is an extra parameter that could be tuned.
Such a mechanism was proposed in \cite{krasnikov}. For stabilizing $S$, the 
proposal was to consider gaugino condensation of a nonsemisimple
 gauge group, inducing a sum of exponentials in the superpotential 
$W(S)\sim \sum_i{\alpha_i \exp(-3S/8\pi b_i)}$
which can conspire to generate a local minimum for $S$ \cite{krasnikov}.
The role of the extra parameter can be played by the ratio of beta
function coefficients of the different groups.
These have been named `racetrack' models in the recent literature.

It was later found that combining the previous ideas, together with 
the addition of  matter fields in the hidden sector (natural in many string models)\cite{lt,ccm},  was sufficient to find a minimum with almost all
the right properties, namely, $S$ and $T$ fixed at the desired value,
$S_R\sim 25, T_R\sim 1$, supersymmetry broken at a small
scale ($\sim 10^{2-4}$ GeV) in the observable sector, etc.
This lead to studies of the induced soft breaking terms at low energies.
Besides that relative success there are several problems that assure us
that we are far from a satisfactory description of these issues
\footnote{Another important puzzle was: we know that the
field $S$  only appears after performing a duality transformation
changing the stringy field $B_{\mu\nu}$ to the axion $a$. A non-trivial potential for $S$ gives a mass to $a$ and then it is no longer dual to
$B_{\mu\nu}$! This puzzle was recently solved
\cite{bdqq2}\ by analyzing gaugino
condensation directly in the $B_{\mu\nu}$ version. The end result was that
$B_{\mu\nu}$ dissapears from the low-energy spectrum and a {\em massive} $H_{\mu\nu\rho}$ field takes its place, having one 
propagating degree of freedom and being   dual to a massive axion $a$.}

\begin{description}
\item[(i)] Unlike the case for $T$, fixing  the
$vev$ of the dilaton field $S$, at the phenomenologically
interesting value, is not achieved in a satisfactory way.
The conspiracy of several condensates with hidden
matter to generate a local minimum
at a good value, requires certain amount of fine tunning and
cannot be called natural. One of the motivations for proposing
the existence of a $S$-duality was precisely to find a
 way in which the vev of $S$ could be fixed in a natural way as it happens for $T$ \cite{filq2}. More recently, there have been attempts to combine gaugino condensation with $S$ duality
\cite{hm}, in which $S$ can be fixed at a 
selfdual point $S=i, e^{i\pi/6}$; but, besides the present
ignorance of how $S$-duality can be realized in 4D effective actions,
 there is also an unjustified assumption that 
this effect will provide the dominant non-perturbative correction to the
superpotential (see for instance \cite{mirev}\, for 
a discussion of these points). Possible arguments for this
to be the case were given in \cite{bdine}. There, it was also  proposed 
to
use 
 the recently found non-perturbative behaviour of string models, of the form
$e^{-1/g}$ \cite{joe}\ (rather than the field theoretical $e^{-1/g^2}$).
These corrections in the K\"ahler potential can in principle combine
with a superpotential like that of eqs. (42,43), to fix the value of $S$.
There is no yet a concrete case where these ideas are  realized, though.

\item[(ii)] The  cosmological constant turns out to be
always negative,
which looks like an unsourmountable problem at present. This
also makes the analysis of soft breaking terms less reliable,
because in order to talk about them, a constant 
piece has to be added to the Lagrangian   that 
cancels the cosmological constant. It is then hard to 
believe that the unknown mechanism generating this term would
leave the results on  soft breaking terms (such as the
smallness of gaugino masses) untouched.

\item[(iii)] Finally, even if the previous problems were solved, there are
at least two serious cosmological problems for the gaugino condensation
scenario. First, it was found under very general grounds, that 
it was not possible to get inflation with the type of dilaton potentials
obtained from gaugino condensation \cite{bs}. Second is the so-called
`cosmological moduli problem' which applies to any (non-renormalizble)
hidden sector  scenario including gaugino condensation
\cite{modprob, dcrq}. In this case,
it can be shown that if the same effect that 
fixes the vev's of the moduli, also breaks supersymmetry,
then: the moduli and dilaton fields acquire masses
of the electroweak scale ($\sim 10^2$ GeV) after supersymmetry breaking
\cite{dcrq}.
Therefore if  stable, they overclose the universe, if
 unstable, they  destroy  nucleosynthesis by their 
late decay, since they only have gravitational strength interactions.
At present there is no satisfactory explanation of this problem and it stands
as one of the unsolved generic problems of string phenomenology.

\end{description}

% {Wilson vs 2PI Actions}

\vspace{4.0 mm}
{\it \noindent Soft SUSY Breaking Terms}
\vglue 0.4cm

It was found many years ago that some terms can be added
to a supersymmetric lagrangian that do not respect
supersymmetry but still keep the soft ultraviolet 
behaviour of the theory. These are the soft-breaking terms. They
are naturally generated after supersymmetry breaking in general
supergravity models and correspond to the following terms
\footnote{On top of this soft terms, we have to consider the induced quadratic divergences,
for a recent discussion of this see \cite{qsquare}.}:

\begin{description}

\item{(i)} Scalar masses, implying that the scalars such as squarks, 
become usually heavier than the fermions of the same multiplet.
These are terms in the Lagrangian of the form $m_I^2\left|Q^I\right|^2$.

\item{(ii)} Gaugino masses $M_a\lambda^a\lambda^a$ splitting the gauge
multiplets.

\item{(iii)}Cubic $A$ terms. Cubic scalar terms in the potential related
to Yukawa couplings and controlled by arbitrary dimensionfull
coefficients ($A$) of order the gravitino mass.

\item{(iv)}The $B$-term. A quadratic term in the potential for the scalars
of the form $B\mu H\ov H$ where $H, \ov H$ represent the Higgs fields and
$\mu$ is a constant that gives rise to a term in the original
superpotential $W=\mu H\ov H$ which is allowed by all the
symmetries of the minimal supersymmetric standard model.
Since $\mu$ is a dimensionfull parameter it causes a problem to introduce it 
in the supersymmetric Lagrangian since it has to be of the order of
the gravitino mass and there is no reason that a term in the
supersymmetric lagrangian knows the scale of the breaking of
supersymmetry.This is known as the $\mu$ problem and several
solutions have been proposed. Depending on the proposed solution there is
an expression for the parameter $B$ after supersymmetry breaking.
In particular, if $Z\neq 0$ in eq. (21), it can be seen that 
the $\mu$ term is generated after supersymmetry breaking. Some calculations
have shown that there are models for which $Z\neq 0$ (for recent discussions see \cite{at,carlos}\ ). 

\end{description}

In this section we will follow the following strategy.
Treat the supersymmetry breaking mechanism as a black box,
but based on the experience with gaugino condensation,
use that the end result of this mechanism is to induce nonvanishing
values to the auxiliary fields of the moduli or dilaton fields.
Therefore we can parametrize our ignorance of the 
particular breaking mechanism by working with general values of these auxiliary fields. Let us for simplicity treat a single modulus field
$T$ and the dilaton $S$. But the analysis has been done in more general
cases \cite{cfilq,il,kln,bin}. We can then write the goldstino field (Goldstone fermion
eaten by the gravitino in the process of supersymmetry breaking) as a linear combination of the fermionic components of $S$ and $T$
\cite{bin}:
\beq
\tilde \eta=\sin\theta\, \tilde S+\cos\theta\, \tilde T
\eeq
where the goldstino angle $\theta$ mixing $\tilde S$ and $\tilde T$
describes the relative contribution of 
$S$ and $T$ to the breaking of supersymmetry.
The general procedure for extracting the soft breaking terms is clear.
We start with the supersymmetric lagrangian and substitute  in it the nonvanishing auxiliary fields
($\sim e^{K/M_p^2} K^{-1/2}_{z\ov z}D_zW$), using expression (44). Performing the so called flat space limit in which $M_p\rightarrow \infty$
with fixed gravitino mass $m_{3/2}$ (representing the nonvanishing vev's of the auxiliary fields and parametrizing the breaking of supersymmetry.
 We end up with the following values for the soft breaking parameters
\cite{bin}:
\beqa
m_I^2 &=&m_{3/2}^2 \left(1+n_I\cos^2\theta\right)\nonumber\\
M^a&=&\sqrt 3 m_{3/2}\sin\theta\nonumber\\
A_{IJK}&=&-\sqrt{3} m_{3/2}\left(\sin\theta+\cos\theta (n_I+n_J+n_K)\right)
\eeqa
From this we can extract several conclusions. The dilaton dominated scenario
 for which $\sin\theta=1$, the soft breaking parameters are
{\it universal}!This is a very appealing result explaining one of the
less justified assumptions of the minimal supersymmetric standard model.
On the other hand, this scenario is so restrictive that it is
relatively easy to rule it out. Something that  was claimed recently
after comparison with the value of the top quark mass, using a particular solution of the
$\mu$ problem above  (see for instance
\cite{carlos} for a recent discussion). The importance of this scenario is that 
eqs (45) for $\sin\theta=1$ are valid in general and not only for
orbifold models.
For arbitrary mixing angle, the soft breaking terms are not 
necessarily universal (unless special values of the modular weights
are taken). In that case we have to confront problems with flavour changing neutral currents \cite{bin}.
Another conclusion we can extract from the form of the soft breaking terms is that 
for negative modular weights, we can get tachyons in the spectrum,
for certain values of the mixing angle. The same condition that avoids tachyons, implies that the gaugino masses have to be bigger than
the scalar masses, unless both vanish; in that case loop corrections may be important to determine the relative masses.    

 \vspace{4.0 mm}
{\it \noindent Scenarios for SUSY Breaking}
\vglue 0.4cm

The results of the previous sections have shown us that 
the general results extracted in the past years about
gaugino condensation in string models, in terms of the
field $S$,  are robust.
We have seen how gaugino condensation can in principle 
lift the string vacuum degeneracy and break supersymmetry 
at low energies (modulo the problems mentioned before).
But this is a very particular field theoretical mechanism
and it would be surprising that other nonperturbative effects
at the Planck scale could be completely irrelevant for these
issues. In general we should always consider the two types
of nonperturbative effects:stringy (at the Planck scale) and
field theoretical (like gaugino condensation). Four different scenarios
can be considered depending on which class of mechanism solves
each of the two problems:lifting the vacuum degeneracy and breaking
supersymmetry. 

 For breaking supersymmetry at low energies, we expect that
a field theoretical effect should be dominant in order
to generate the hierarchy of scales (it is hard to believe that
a nonperturbative effect at the Planck scale could generate
the Weinberg-Salam scale). 
We are then left with two preferred scenarios:
either the dominant nonperturbative effects are field theoretical,
solving both problems simultaneously, or there is a `two steps' scenario 
in which stringy effects dominate to lift  vacuum degeneracy 
and field theory effects dominate to break supersymmetry.
The first scenario has been the only one considered so far,
it includes gaugino condensation and also the discussion of the
previous section in terms of field-dependent soft breaking terms.
The main reason this is the only scenario considered so far is that we can control field theoretical
nonperturbative effects but not the stringy. In this scenario,
 independent of the particular mechanism, we have to face the cosmological moduli problem.

In the two steps scenario 
the dilaton and moduli fields are fixed at high energies
with a mass $\sim M_{Planck}$ thus avoiding the cosmological moduli problem.
It is also reasonable to expect that Planck scale
effects can generate a potential for $S$ and $T$. 
The problem resides in the implementaion of this scenario
\cite{biq, mirev}, 
mainly due
to our ignorance of nonperturbative string effects.

%{\it Two Steps Scenario}

In the two steps scenario, after we have fixed the vev
of the moduli by stringy effects, it remains the question of
how supersymmetry is broken at low energies. Notice that we
would be left with the situation present before the advent
of string theory in which the gauge coupling is
 field {\it independent}. In that 
case we know from Witten's index that gaugino condensation cannot 
break global supersymmetry. Since there are no `moduli'
fields with large vev's, the supergravity correction should
be negligible because we are working at energies much smaller
than $M_{Planck}$.

 In fact we can perform a calculation by
setting $S$ to a constant in eq.~(41),
 it is straightforward to show that supersymmetry
is still unbroken in that case \cite{biq}, as expected. A more general  
way to see this is
computing explicitly the $1/M_{Planck}$ correction to a
global supersymmetric solution $W_\phi=0$, and see that it
coincides with the solution of 
$W_\phi+WK_\phi/M_p^2=0$ which  is always a 
supersymmetric extremum of 
the supergravity scalar potential.

There seems to be however a counterexample in the literature
\cite{peter},
where supersymmetry was found to be  broken with vanishing cosmological constant in supergravity but unbroken in global supersymmetry.
Nevertheless it can be seen that  
in that case, the global limit is such that $K_{UU^*}$ vanishes, and so the kinetic energy
for $U$. This makes the corresponding
 minimum in the global case ill defined, since there may be other 
 nonconstant field configurations with vanishing energy.
This is then not a counterexample, because the
global theory is not well defined in the minimum.

We are then left with a situation that if global supersymmetry
is unbroken, we cannot break local supersymmetry, unless
there are moduli like fields. 
If we insist to have the two steps scenario, 
this can bring us further back to the past and reconsider models 
with dynamical
breaking of global supersymmetry. These models are not only very complicated but have some phenomenological problems that need
to be addressed before 
they can be taken as valid alternatives, in particular, issues related with small
gaugino masses and the value of the $\mu$ parameter are not 
naturally solved in these models. Nevertheless, in view of the cosmological moduli problem and because of the recent progress in understanding $N=1$ field theories, these
models are attracting attention recently and, hopefully, some progress
will be made
(for a recent discussion with new insights see \cite{dnn}\ ).

Therefore, there is not yet a compelling
 scenario for supersymmetry breaking
and the field remains open, but now we have a much 
better perspective on the relevant issues. The
nonrenormalizable hidden sector models of which the 
gaugino condensation is
a particular case, may need a convincing  solution of the
cosmological moduli problem to still be considered 
viable. Hopefully, this will lead to interesting feedback
between cosmology and string theory \cite{bms}
\footnote{ A good example of this string-cosmology interaction is
the recent work by the authors of ref.\cite{bgv}, on which
investigations on string cosmology is leading to interesting
{\it experimental} searches for gravitational waves in ranges not explored before.}.
Furthermore, the recent progress in understanding 
supersymmetric gauge theories can be of much use for 
reconsidering gaugino condensation with hidden matter,
the discussion in the string literature is far from complete. 
The understanding of models with chiral 
matter could also provide new insights to 
global supersymmetry
breaking, relevant to the
two steps scenario mentioned above. 

\vspace{4.0 mm}
{\large\bf\noindent Outlook}
\vglue 0.4cm

We have seen that the field of string phenomenology has been very 
animated
during the past ten years and that a lot of progress has been achieved.
There is a very  natural justification for this kind of research, namely
to 
look for  a connection between string theory and physics. 
Nevertheless we have seen that this field has also contributed enormously
to the progress towards formulating string theory itself. 
It has been claimed that string theory is currently in  a similar
situation as particle physics was in the early sixties, in the sense that
there are a lot of experiments without a theory \cite{joe}.
The only difference is that in string theory, the `experiments'
are just the data we have from string model building.
These data have been fundamental in the discovery of  $T$-duality and mirror
symmetry in string theory, furthermore, the conjectured
$S$-duality \cite{filq2}, which relates strong to weak coupling in string theory,
was also originated by asking phenomenological questions.
A symmetry of this sort was also fundamental to exactly {\it solve}
$N=2$ field theories, even at the non-perturbative level, in the
impressive work of Seiberg and Witten  that  lead to the
understanding of confinement in a non-trivial 4D field theory\cite{sw}. 
Even though that result
was independent of string theory, the full mathematical treatment was
inherited from recent progress in string compactifications, in particular
the study of mirror symmetries and $T$-duality in Calabi-Yau 
compactifications.

We mentioned in the introduction that there are five consistent superstring theories and each has thousands or millions of different vacua.
It is now believed that the five string theories are related by 
strong-weak coupling dualities and furthermore, they appear 
to be different limits of a single underlying fundamental theory,
probably in 11D, the $M$ theory
(probably related with membranes or higher dimensional objects such as five-branes), 
which is yet to be constructed.
If this is true it may solve the arbitrariness in the number 
of fundamental string theories by deriving them from a single
theory\footnote{We may hope that this recent progress will lead to 
the answer of the `why' question mentioned in the introduction and select 
 our 
Universe uniquely from the underlying fundamental theory.
Otherwise, we might have to invoke the anthropic principle and probably
 imagine our Universe emerging in a kind of darwinian natural selection. Many theorists disregard this second option because it implies that the theory would not have predictive power.
However, this posture may be too naive, it is only a `philosophical prejudice'
similar to the geocentric ideas of   Aristotle since it is like assuming that our Universe has to be the  only 
possible outcome of a fundamental theory. Regardless of any philosophical prejudice, we  have to study the theory to its limits. This is the attitude that string phenomenologists have
been taking during the past 10 years.
We may hope that, if there is a fundamental theory and
still many possible models are allowed, there could still be several general features that are model independent, such as those mentioned for QFT in the introduction. They could help in  eventually testing the theory and not only the particular models.}. Furthermore, recent work
based on comparison of string compactifications with the Seiberg-Witten
theory, has lead to the conclusion that many and probably all
Calabi-Yau compactifications are connected. Then it seems
that not only the five different theories are unified, but also
all the vacua of these theories are unified: since, if they are all connected,  we can foresee a mechanism that lifts the
degeneracy and select one point in the web
of compactifications, something it could not have been done before 
because they were thought to be disconnected vacua and there
was no way to select one. Also the Seiberg-Witten
results were reproduced from a field theory approximation of 
a particular string vacuum, assuming $S$ duality to hold.
Since the Seiberg-Witten results are  robust, this is a 
 non-trivial test for the existence of $S$ duality in string 
theories (for a recent review see \cite{joe}\ ).

We can see that we are living  exciting times in string theory,
where not only strings but also higher-dimensional objects are emerging as different pieces of an underlying fundamental theory.
In that sense we can repeat the famous phrase of Weinberg
\cite{steve2}:
{\em The more  the Universe seems comprehensible, the more it seems
pointless}; but in a less nostalgic fashion, claiming that the universe looks point-less but full of
higher-dimensional objects such as strings! On the other hand 
the current optimism for this theory should be taken in the
right perspective, none of these achievements can 
substitute a solid experimental test of the theory, something
which may still be very far in the future. However we might be
lucky and some discoveries at current and future accelerators,
as well as some possible astrophysical results,  could provide 
important clues on the validity of string theory.

\vspace{4.0 mm}
{\large\bf\noindent Acknowledgements \hfil}
\normalsize
\vglue 0.4cm

I thank the organizers of the conference for providing the opportunity to
give these lectures. I would also like to acknowledge my 
collaborators during the last few years who have influenced
my understanding of this subject. Finally I would like to
thank useful conversations with L.E. Ib\'a\~nez,
S.-J. Rey and J. Russo that influenced the way the
subjects were presented.

\vspace{4.0 mm}
{\large\bf\noindent References \hfil}
\normalsize
\vglue 0.4cm

\end{document}